\documentclass[11pt,draftcls,onecolumn]{IEEEtran}
\usepackage{caption2}
\usepackage{graphicx}
\usepackage{subfigure}
\usepackage{booktabs}
\usepackage{multirow}
\usepackage{cite}%
\usepackage[
            linkcolor=red,
            anchorcolor=blue,
            citecolor=green]{hyperref}

\usepackage{amsmath}
\usepackage{amssymb}
\usepackage{bm}
 \usepackage{color}
\usepackage{verbatim}

\ifCLASSINFOpdf
\else
\fi

\begin{document}

\title{Colocated MIMO radar waveform design for transmit beampattern formation}

\author{\IEEEauthorblockN{Haisheng Xu\IEEEauthorrefmark{1},
Rick S. Blum\IEEEauthorrefmark{2},~\IEEEmembership{Fellow,~IEEE},
Jian Wang\IEEEauthorrefmark{1},~\IEEEmembership{Member,~IEEE}, and
Jian Yuan\IEEEauthorrefmark{1}
}\\
\IEEEauthorblockA{\IEEEauthorrefmark{1}Department of Electronic Engineering, Tsinghua University, Beijing 100084, P. R. China}\\
\IEEEauthorblockA{\IEEEauthorrefmark{2}Department of Electrical and Computer Engineering, Lehigh University, Bethlehem, PA, 18015, USA}\\

\thanks{This work was done at Lehigh University and partially supported by the National Science Foundation
under grant ECCS-1405579 and by the Tsinghua Short-time Visiting Foundation under grant 2013048.}
\thanks{Authors' addresses: H. Xu, Department of Electronic Engineering,
                Tsinghua University, Beijing 100084, P. R. China, e-mail: xhs11@mails.tsinghua.edu.cn; R. S. Blum, Department of Electrical and Computer Engineering, Lehigh University, Bethlehem, PA 18015, USA,
                e-mail: rblum@eecs.lehigh.edu; J. Wang \& J. Yuan, Department of Electronic Engineering,
                Tsinghua University, Beijing 100084, P. R. China, e-mail: jian-wang\{jyuan\}@tsinghua.edu.cn.}
}

\markboth{ACCEPTED BY IEEE TRANSACTIONS ON AEROSPACE AND ELECTRONIC SYSTEMS}
{Shell \MakeLowercase{\textit{et al.}}: Bare Demo of IEEEtran.cls for Journals}

\IEEEtitleabstractindextext{

\begin{abstract}
  In this paper, colocated MIMO radar waveform design is considered by minimizing the integrated sidelobe level (ISL) to obtain beampatterns with lower sidelobe levels than competing methods. First, a quadratic programming problem is formulated to design beampatterns by using the minimal-ISL criteria. A theorem is derived that provides a closed-form analytical optimal solution that appears to be an extension of the Rayleigh quotient minimization for a possibly singular matrix in the quadratic form. Such singularities are shown to occur in the problem of interest, but proofs for the optimum solution in these singular matrix cases could not be found in the literature. Next, an additional constraint is added to obtain beampatterns with desired 3dB beamwidths, resulting in a nonconvex quadratically constrained quadratic program which is NP-hard. A semi-definite program and a Gaussian randomized semi-definite relaxation are used to determine feasible solutions arbitrary close to the solution to the original problem. Theoretical and numerical analyses illustrate the impacts of changing the number of transmitters and orthogonal waveforms employed in the designs. Numerical comparisons are conducted to evaluate the proposed design approaches.
\end{abstract}

}

\maketitle

\IEEEdisplaynontitleabstractindextext
\IEEEpeerreviewmaketitle

\section{Introduction}
Multi-input multi-output (MIMO) radars which emit different waveforms from different transmit antennas can
provide extra degrees of freedom in waveform design to allow
significant coherent gains \cite{Fuhrmann2004,Li2009,Gorji2012,Cui2014} when the antennas are colocated. Thus, flexible waveform design for transmit beampattern formation is of interest for colocated multi-input multi-output (MIMO) radars. Beampattern formation is carried out by designing either a waveform covariance matrix \cite{Li2006,Li2008a} or a waveform coefficient matrix \cite{Khabbazibasmenj2014}. The goal is to focus the transmit
power into a range of interesting angles while minimizing the transmit power for other angles \cite{Aittomaki2007}.  Various methods can be found in \cite{Fuhrmann2004,Stoica2007,Li2007a,Aittomaki2007a,Li2007,Fuhrmann2008,Stoica2008,
Aittomaki2009,Ahmed2011,Xu2013,Khabbazibasmenj2014} and in the references therein. Among the references on beampattern design, the most common strategy is called shape approximation, i.e., beampattern design according to a known shape under the criteria of minimum mean square error (MMSE) \cite{Fuhrmann2004,Aittomaki2007a,Fuhrmann2008,Stoica2008,Ahmed2011,Stoica2007} or  minimum difference (MD) \cite{Fuhrmann2008,Khabbazibasmenj2014}. To accomplish these goals, gradient search algorithms \cite{Fuhrmann2004,Aittomaki2007a}, barrier methods \cite{Fuhrmann2008}, iterative algorithms \cite{Stoica2008,Ahmed2011}, and convex optimization \cite{Stoica2007,Khabbazibasmenj2014} have all been used. Since the sidelobe level is one of the most important performance indexes in antenna radiation theory \cite{Gerlach1990,Balanis2005,Liang2014}, some minimum sidelobe level design strategies have received attention and this is also the topic of this paper. Published methods are the minimum peak sidelobe level (PSL) method \cite{Li2007,Stoica2007,Gong2014,Shariati2014}, the discrete prolate spheroidal sequences-based design (DPSSD) method \cite{Hassanien2011} and the minimum integrated sidelobe level (ISL) method \cite{Xu2013}. The minimum PSL method constrains the definition of the sidelobe region as the region outside a band of angles twice as wide as the mainlobe, which is a severe restriction that limits flexibility to employ wider mainlobes.  DPSSD used a maximization of the ratio of energy in mainlobe to the total energy to derive a closed-form solution using some approximations. Further, the sidelobe levels obtained by maximizing the ratio of energy in mainlobe to total energy can never be smaller than obtained by maximizing the ratio of energy in mainlobe to the energy in sidelobes \cite{Martinezl1993}.
Alternatively, in this paper a closed-form analytical solution is given for the criterion of minimum ISL and no approximations are employed. To the best of our knowledge, this solution and the proof justifying it, have not appeared in any publications to date.
Our results appear to be an extension of the Rayleigh quotient minimization \cite[Sec. 8.2.3]{Golub2013} for a possibly singular matrix in the quadratic form. Such singularities are shown to occur in the problem of interest, but proofs for the optimum solution in these singular matrix cases could not be found in the literature.
To augment the closed form results, the minimum ISL criterion is next considered with
an additional constraint to require a desired 3dB beamwidth,  This
results in a nonconvex quadratically constrained quadratic program which is NP-hard.
A semi-definite program and a Gaussian randomized semi-definite relaxation is used to determine feasible solutions arbitrary close to the solution to the original problem. Extensive numerical comparisons are conducted to evaluate the proposed design approaches. Additionally, a novel theoretical analysis describes the impact of the number of transmitters and the number of transmit waveforms on the designed beampatterns.
\par Notation: Throughout this paper, we use lowercase italic letters to denote scalars, lowercase and uppercase letters in bold to denote the vectors and the matrices, respectively. Superscripts of $\ast$, $T$ and $H$ represent the conjugate, transpose and complex conjugate-transpose (or Hermitian) operators of a matrix, respectively, while $\mathrm{tr}\{\cdot\}$ represents the trace of a square matrix and $\mathrm{vec}\{\cdot\}$ represents the vectorization operator which creates a column vector from a matrix by stacking its column vectors. $\|\cdot\|$ denotes the Euclidean (or $l^2$) norm of a vector and $\otimes$ denotes Kronecker product. The notation $\mathbf{I}$ denotes identity matrix and $\mathbf{0}$ denotes all-zero vector or matrix. $\mathbf{\mathbb{R}}$, $\mathbf{\mathbb{C}}$ and $\mathbf{\mathbb{Z}}^{+}$ denote the real, complex and positive integer spaces, respectively.
\section{Problem formulation}\label{sec:SMTB}
\par Consider a colocated uniform linear array (ULA) MIMO radar system equipped with $M$ transmit antennas, where the antennas are separated by a half wavelength. Assume each transmitter emits a weighted sum of $Q$ independent orthonormal baseband waveforms which are expressed by $\bm\phi\left( t \right) = \left[ {\phi _1 \left( t \right)}~ {\phi _2 \left( t \right)}~\cdots~{\phi _Q \left( t \right)}\right]^T$. The weightings are described by a coefficient matrix $\mathbf{C}=[\mathbf{c}_{1}~\mathbf{c}_{2}~\cdots~\mathbf{c}_{Q}]$, where $\mathbf{C} \in \mathbf{\mathbb{C}}^{M\times Q}$ and $\mathbf{c}_q$ is used to control the amount of the $q$th waveform added into a sum at each transmit antenna \cite{Hassanien2011}. The total transmit power is fixed at $E$ by setting $\sum\limits_{q = 1}^Q {\left\| {\mathbf{c}_q } \right\|^2 }  = E$. Thus the signals transmitted by the $M$ antennas can be described by
\begin{equation}\label{equ:Tsignal}
  \bm\psi \left( t \right) = \mathbf{C}\bm\phi\left( t \right).
\end{equation}
\par
After applying a steering vector $\mathbf{a}(\theta)$, the waveforms transmitted to a far-field target at $\theta$ with $\theta\in\Theta=[-\pi/2, \pi/2]$ can be expressed as
\begin{equation}\label{equ:signals}
s\left( {t,\theta } \right) = \mathbf{a}^H \left( \theta  \right)\mathbf{C}\bm\phi \left( t \right)
\end{equation}
where $\mathbf{a}(\theta)=[1~ e^{j\pi\sin(\theta)}~\cdots~e^{j\pi(M-1)\sin(\theta)}]^T$.
\par Integrating $|s\left( {t,\theta } \right)|^{2}$ over a time equal to the support of $\bm\phi\left( t \right)$, then the expression
for the beampattern becomes\cite{Fuhrmann2004,Stoica2007}
\begin{equation}\label{equ:beampattern}
 P(\theta)=\mathbf{a}^H \left( \theta  \right)\mathbf{C}\mathbf{C}^H \mathbf{a}\left( \theta  \right). \\
\end{equation}
\par Given the beampattern $P(\theta)$ expressed in (\ref{equ:beampattern}), based on\cite{Neudecker1969}, we can write it as
\begin{equation}\label{equ:BeampatternTranfo}
\begin{split}
P\left( \theta  \right)&=\|\mathbf{C}^H \mathbf{a}\left( \theta  \right)\|^{2}= \|{{\mathrm{vec}}\left\{ \mathbf{a}^H \left( \theta  \right)\mathbf{C} \right\}}\|^{2}\\
 &=\left\|\left(\left({\mathbf{I}_Q}\otimes{\mathbf{a}^{H}\left( \theta  \right)}\right)\mathrm{vec}\left\{\mathbf{C} \right\}\right)\right\|^{2}\\
 &={\mathrm{vec}^H\left\{ \mathbf{C} \right\}}\left( {\mathbf{I}_Q}\otimes \left( {\mathbf{a}\left(\theta  \right)}{\mathbf{a}^{H}\left(\theta\right)}\right) \right){\mathrm{vec}\left\{ {{\mathbf{C}}} \right\}}\\
 &=\mathbf{c}^{H}\left( {\mathbf{I}_Q}\otimes {\mathbf{A}(\theta)} \right)\mathbf{c}=\mathbf{c}^{H}{\widetilde{\mathbf{A}}(\theta)}\mathbf{c}
\end{split}
\end{equation}
where $\mathbf{c}\triangleq{\mathrm{vec}\left\{ \mathbf{C} \right\}}$, $\mathbf{A}(\theta)\triangleq {\mathbf{a}\left(\theta  \right)}{\mathbf{a}^{H}\left(\theta\right)}$ and $\widetilde{\mathbf{A}}(\theta)\triangleq { {\mathbf{I}_Q}\otimes \mathbf{A}(\theta)} $. Note that both $\mathbf{A}(\theta)$ and $\widetilde{\mathbf{A}}(\theta)$ are Hermitian positive semidefinite matrices for $\forall \theta$.
\par Divide $\Theta$, the set of all possible $\theta$, into two disjoint sets called $\Theta_{ml}$, the mainlobe region and $\Theta_{sl}$, the sidelobe region. To concentrate the transmit power into $\Theta_{ml}$ as much as possible, we can formulate the following minimum integrated sidelobe level (ISL) optimization
\begin{equation}\label{equ:optimization}
\begin{split}
\min\limits_{\|\mathbf{c}\|>0}&~~\frac{{{\mathbf{c}^H}{\mathbf{A}_{sl}}\mathbf{c}}}{{{\mathbf{c}^H}{\mathbf{A}_{ml}}\mathbf{c}}}\\
\mathrm{s.t.}&~~{\left\| \mathbf{c} \right\|^2} = E
\end{split}
\end{equation}
where
\begin{equation}\label{equ:factor1}
\mathbf{A}_{sl}\triangleq\int_{\Theta_{sl}} {\widetilde{\mathbf{A}}\left( \theta  \right)d\theta },~~~\mathbf{A}_{ml} \triangleq\int_{\Theta_{ml}} {\widetilde{\mathbf{A}}\left( \theta  \right)d\theta}.
\end{equation}
\par
\begin{figure}[ht]
  {\begin{minipage}{1\textwidth}
  \centering
  \includegraphics[scale=0.95]{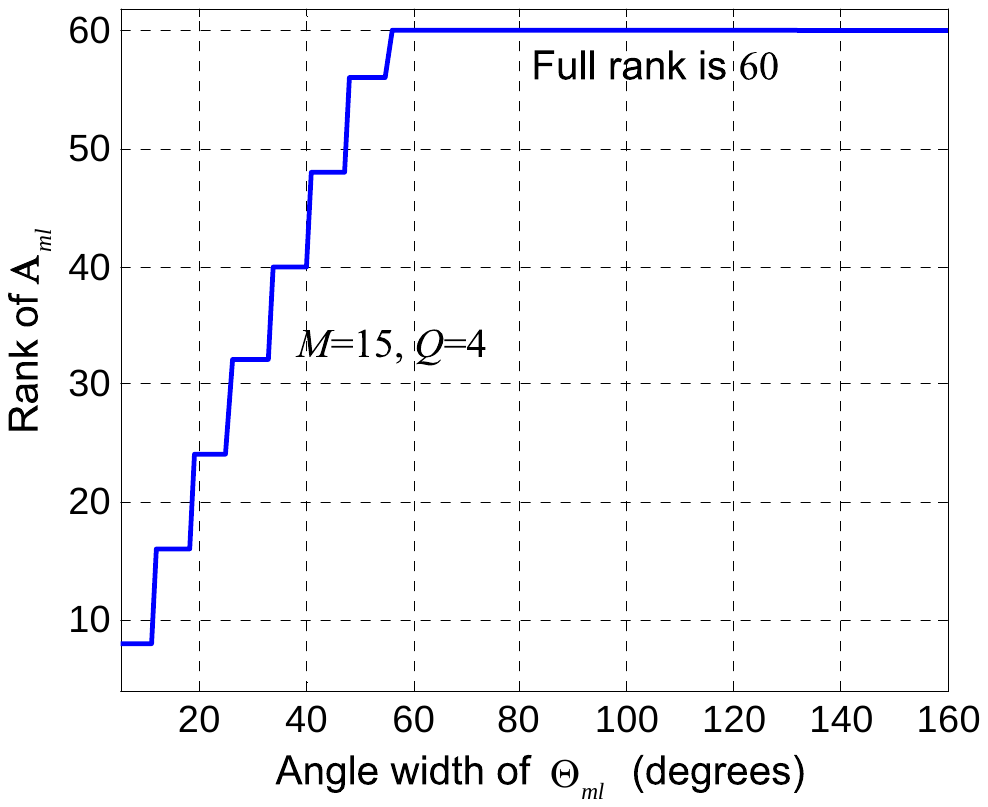}
  \end{minipage}}\\
  \caption{Demonstration on the rank of $\mathbf{A}_{ml}$ vs. angle width of $\Theta_{ml}$}\label{fig:1}
\end{figure}

 Problem (\ref{equ:optimization}) is a quadratic programming problem with a ratio objective (QP-R).
When $\mathbf{A}_{ml}$ ($\in \mathbf{\mathbb{C}}^{MQ\times MQ}$) is a positive definite matrix with full
rank, then (\ref{equ:optimization}) just can be converted into a Rayleigh quotient \cite[Sec. 8.2.3]{Golub2013} minimization,
and an analytical optimal solution can be obtained \cite{Beck2009,Cai2014,Beck2010}. However, in our case $\mathbf{A}_{ml}$ may not be a full rank matrix which means $\mathbf{A}_{ml}$ will not be positive definite. Here we provide a numerical demonstration of this by considering a case when we vary the width of mainlobe from $5^{\circ}$ to $160^{\circ}$ in Fig. \ref{fig:1}. From the figure, we can see that only when the width of
 $\Theta_{ml}$ is greater than $55^{\circ}$ is $\mathbf{A}_{ml}$ full rank. Thus, a more general method to obtain the analytical optimal solution of (\ref{equ:optimization}) needs to be considered. However, to the best of our knowledge, we have not found any literature giving an analytical solution to (\ref{equ:optimization}) for $\mathbf{A}_{ml}$ not full rank.
\par The problem in (\ref{equ:optimization}) seeks the minimal ISL for the defined $\Theta_{ml}$ and $\Theta_{sl}$.  However, in some applications a particular $3$dB beamwidth is also desired, requiring the addition of some constraints to (\ref{equ:optimization}) to obtain the altered optimization by
\begin{equation}\label{equ:optimizationBW}
\begin{split}
\min\limits_{\mathbf{c}}&~~\frac{{{\mathbf{c}^H}{\mathbf{A}_{sl}}\mathbf{c}}}{{{\mathbf{c}^H}{\mathbf{A}_{ml}}\mathbf{c}}}\\
\mathrm{s.t.}&~~\frac{1}{2}\leq\frac{\mathbf{c}^H{\widetilde{\mathbf{A}}(\theta)}\mathbf{c}}{\mathbf{c}^H{\widetilde{\mathbf{A}}(\theta_{0})}\mathbf{c}}\leq1,~~\forall\theta\in\Theta_{ml}.\\
\end{split}
\end{equation}
where $\theta_{0}$ denotes the maximum power point in $\Theta_{ml}$, which is usually chosen to be at the center point of the main lobe. The above problem is a nonconvex quadratically
constrained quadratic programming (QCQP) problem \cite{Boyd2004} which is NP-hard.
\par In the following section, we will propose the best suitable methods to solve the above two problems, respectively.
\section{Waveform desgin for Beampattern formation}\label{sec:BPdesign}
\subsection{Beampattern formation for minimal-ISL only}\label{sec:SubSecBPDM}
\par Although we have demonstrated that $\mathbf{A}_{ml}$ might not be a positive definite matrix, we can confirm that $\mathbf{A}_{ml}+\mathbf{A}_{sl}$ is positive definite as shown in the following Lemma 1.
\par \textbf{Lemma 1:} Given $\mathbf{A}_{sl}=\int_{\Theta_{sl}} {{\mathbf{I}_{Q}\otimes{\mathbf{A}(\theta)}}~d\theta}$ and $\mathbf{A}_{ml}=\int_{\Theta_{ml}} {{\mathbf{I}_{Q}\otimes{\mathbf{A}(\theta)}}~d\theta}$, where $\Theta_{sl}+\Theta_{ml}=[-\pi/2,\pi/2]$, $\mathbf{A}(\theta)=\mathbf{a}(\theta)\mathbf{a}^{H}(\theta)$ and $\mathbf{a}(\theta)=[1~ e^{j\pi\sin(\theta)}~\cdots~e^{j\pi(M-1)\sin(\theta)}]^T$, then $\mathbf{A}_{ml}+\mathbf{A}_{sl}$ is always positive definite.\\
\textbf{Proof of Lemma 1: }
\par For ${\mathbf{I}_{Q}\otimes{\mathbf{A}(\theta)}}$ is positive semidefinite, thus given $\forall\mathbf{c}\in \mathbf{\mathbb{C}}^{MQ\times 1}$ and $\|\mathbf{c}\|>0$, we have
\begin{equation}\label{equ:DefiniteProof}
\begin{split}
  \mathbf{c}^{H}\left({\mathbf{A}_{sl}}+{\mathbf{A}_{ml}}\right)\mathbf{c} &=\mathbf{c}^{H}\left(\int_{\frac{-\pi}{2}}^{\frac{\pi}{2}}{{\mathbf{I}_{Q}\otimes{\mathbf{A}(\theta)}}}d\theta\right) \mathbf{c}\\
  &\geq\mathbf{c}^{H}\left(\int_{\frac{-\pi}{2}}^{\frac{\pi}{2}}{\left({{\mathbf{I}_{Q}\otimes{\mathbf{A}(\theta)}}}\right)\cos(\theta)}d\theta\right)\mathbf{c}\\
   &=\mathbf{c}^{H}\left(\mathbf{I}_{Q}\otimes\int_{\frac{-\pi}{2}}^{\frac{\pi}{2}}{{\mathbf{A}(\theta)}{\cos(\theta)}}d\theta\right)\mathbf{c}\\
   &=\mathbf{c}^{H}\left(\mathbf{I}_{Q}\otimes(2\mathbf{I}_{M})\right)\mathbf{c}=2\|\mathbf{c}\|^{2}>0
 \end{split}
\end{equation}
\par Obviously, $\mathbf{A}_{sl}+\mathbf{A}_{ml}$ is always positive definite, hence, the proof of Lemma 1 is completed.$\hfill\blacksquare$
\par Based on Lemma 1, then we can use the following theorem (which can be viewed as an extension of the Rayleigh quotient minimization) to get the minimal solution for problem (\ref{equ:optimization})(Note that $\mathbf{c}$ can normalized without loss of any generality for the ratio objective function, thus the constraint $\|\mathbf{c}\|^{2}=E$ in (\ref{equ:optimization}) is ignored).
\par \textbf{Theorem 1:} Define Hermitian matrices $\mathbf{A},~\mathbf{B}\in \mathbf{\mathbb{C}}^{n\times n}$, and assume $\mathbf{A}+\mathbf{B}$ is positive definite while $\mathbf{A}$ is non-negative definite with $\mathrm{rank}(\mathbf{A})=\varsigma~(0<\varsigma\leq n)$. Focusing on $\mathbf{r}\in \mathbf{\mathbb{C}}^{n\times 1}$ and a minimization problem formulated as
\begin{equation}\label{equ:theorem1}
\begin{split}
\min\limits_{\|\mathbf{r}\|>0}&~~\frac{{{\mathbf{r}^H}{\mathbf{B}}\mathbf{r}}}{{{\mathbf{r}^H}{\mathbf{A}}\mathbf{r}}}\\
\end{split}
\end{equation}
then
 \begin{enumerate}
 \item If $0<\varsigma<n$, the minimal solution of (\ref{equ:theorem1}) is given by
 \begin{equation}\label{euq:theorem1c}
  \mathbf{r}=\mathbf{U}\left( {\begin{array}{*{20}{c}}
\left(\bm\Lambda_{1}^{-\frac{1}{2}}\right)^H\\
{ - {\widetilde{\mathbf{B}}_{1}^{- 1}}\widetilde{\mathbf{B}}_{2}^{H} \left(\bm\Lambda_{1}^{ - \frac{1}{2}}\right)^H}
\end{array}} \right)\mathbf{x}_{min}
\end{equation}
where $\mathbf{x}_{min}$ ($\in \mathbf{\mathbb{C}}^{\varsigma\times 1}$) denotes the eigenvector corresponding to the minimum eigenvalue of
${\bm\Lambda_{1}^{-\frac{1}{2}}}\widetilde{\mathbf{B}}{\left(\bm\Lambda _{1}^{-\frac{1}{2}}\right)^H}$; $\widetilde{\mathbf{B}}\in\mathbf{\mathbb{C}}^{\varsigma\times\varsigma}$
such that $\widetilde{\mathbf{B}}\triangleq \widetilde{\mathbf{B}}_{0}-{{\widetilde{\mathbf{B}}}_{2}}{{\widetilde{\mathbf{B}}}_{1}^{-1}} {{\widetilde{\mathbf{B}}}_{2}^{H}}$;
$\bm\Lambda_{1}\in\mathbf{\mathbb{R}}^{\varsigma\times\varsigma}$ is an invertible diagonal matrix obtained from the
eigen-decomposition of $\mathbf{A}$ by $\mathbf{A}=\mathbf{U}\left(\begin{array}{cc}
                                                   \bm\Lambda_{1} & \mathbf{0} \\
                                                   \mathbf{0} & \mathbf{0}
                                                 \end{array}
\right)\mathbf{U}^{H}$, and $\widetilde{\mathbf{B}}_{0}\in\mathbf{\mathbb{C}}^{\varsigma\times\varsigma}$, $\widetilde{\mathbf{B}}_{1}\in\mathbf{\mathbb{C}}^{(n-\varsigma)\times(n-\varsigma)}$ and $\widetilde{\mathbf{B}}_{2}\in\mathbf{\mathbb{C}}^{\varsigma\times(n-\varsigma)}$ are the submatrices of $\mathbf{U}^{H}\mathbf{B}\mathbf{U}$ given by
\begin{equation}\label{equ:theorem1b}
   \mathbf{U}^{H}\mathbf{B}\mathbf{U}=\left(\begin{array}{cc}
                                         \widetilde{\mathbf{B}}_{0} & \widetilde{\mathbf{B}}_{2} \\
                                         \widetilde{\mathbf{B}}_{2}^{H} & \widetilde{\mathbf{B}}_{1}
                                       \end{array}
              \right).
\end{equation}
       \item If $\varsigma=n$, the minimal solution of (\ref{equ:theorem1}) is given by
        \begin{equation}\label{equ:theorem2c}
   \mathbf{r}=\left(\mathbf{A}^{-1/2}\right)^{H}\mathbf{x}_{min}
\end{equation}
where $\mathbf{x}_{min}$ ($\in\mathbf{\mathbb{C}}^{n\times 1}$) denotes the eigenvector corresponding to the minimum eigenvalue of $\mathbf{A}^{-1/2}\mathbf{B}(\mathbf{A}^{-1/2})^{H}$.
     \end{enumerate}
\textbf{Proof of Theorem 1:}\\
1) {$0<\varsigma<n$}
\par $\mathbf{A}$ is a Hermitian and non-negative matrix with rank$(\mathbf{A})=\varsigma$ by assumption. Using eigen-decomposition, we can factorize ${{\mathbf{r}^H}{\mathbf{A}}\mathbf{r}}$ as
\begin{equation}\label{equ:AppendixA3}
\begin{split}
  {\mathbf{r}^H}{\mathbf{A}}\mathbf{r}&=\mathbf{r}^{H}\mathbf{U}\left(\begin{array}{cc}
                     \bm\Lambda_{1} & \mathbf{0} \\
                     \mathbf{0} & \mathbf{0 }
                   \end{array}\right)\mathbf{U}^{H}\mathbf{r}\\
                   &=\mathbf{y}^{H}\left(\begin{array}{cc}
                     \bm\Lambda_{1} & \mathbf{0} \\
                     \mathbf{0} & \mathbf{0 }
                   \end{array}\right)\mathbf{y}\\
                                        &=\mathbf{y_{1}}^{H}\bm\Lambda_{1}\mathbf{y_{1}}
\end{split}
\end{equation}
where $\mathbf{U}$ is a unitary matrix whose columns are eigenvectors argumented with a set of vectors to form a basis and $\bm\Lambda_{1}\in \mathbf{\mathbb{R}}^{\varsigma\times\varsigma}$ is an invertible diagonal matrix with positive eigenvalues of $\mathbf{A}$ along the diagonal. Further, $\mathbf{y}$ is defined as
\begin{equation}\label{equ:AppendixA4}
  \mathbf{y}=\mathbf{U}^{H}\mathbf{r}=\left(\begin{array}{c}
                               \mathbf{y}_{1} \\
                               \mathbf{y}_{2}
                             \end{array}
  \right)=\left(\begin{array}{c}
            \mathbf{U}_{1}^{H}\mathbf{r} \\
            \mathbf{U}_{2}^{H}\mathbf{r}
          \end{array}\right).
\end{equation}
\par Then (\ref{equ:theorem1}) can be formulated as
\begin{equation}\label{equ:AppendixA5}
\min\limits_{\|\mathbf{r}\|>0}~~\frac{{{\mathbf{r}^H}{\mathbf{B}}\mathbf{r}}}{{{\mathbf{r}^H}{\mathbf{A}}\mathbf{r}}}
=\min\limits_{\|\mathbf{y_{1}}\|>0}~~\frac{\mathbf{y}^{H}\mathbf{U}^{H}{\mathbf{B}}\mathbf{U}\mathbf{y}}{\mathbf{y_{1}}^{H}{{\bm\Pi}{\bm\Pi^{H}}}\mathbf{y_{1}}}
\end{equation}
where $\bm\Pi$ is invertible and $\bm\Pi\triangleq\bm\Lambda_{1}^{1/2}$ for $\bm\Lambda_{1}={\bm\Pi}{\bm\Pi^{H}}$.
\par Let $\mathbf{x}=\bm\Pi^{H}\mathbf{y}_{1}$. Further simplification
of (\ref{equ:AppendixA5}) leads to
\begin{equation}\label{equ:AppendixA6}
\begin{split}
\min\limits_{\|\mathbf{r}\|>0}~~\frac{{{\mathbf{r}^H}{\mathbf{B}}\mathbf{r}}}{{{\mathbf{r}^H}{\mathbf{A}}\mathbf{r}}}
&=\min\limits_{\|\mathbf{x}\|>0}~~\frac{\left(\begin{array}{cc}
                                           \mathbf{y}_{1}^{H}~\mathbf{y}_{2}^{H}
                                         \end{array}\right)
                                         \mathbf{U}^H{\mathbf{B}}\mathbf{U}\left(\begin{array}{c}
                                           \mathbf{y}_{1} \\
                                           \mathbf{y}_{2}
                                         \end{array}
\right)}{{\mathbf{x}^H}\mathbf{x}}\\
&=\min\limits_{\|\mathbf{x}\|>0}~~\frac{\left(\begin{array}{cc}
                                           \mathbf{x}^{H}~\mathbf{y}_{2}^{H}
                                         \end{array}\right)
                                         \bm\Sigma\left(\begin{array}{c}
                                                        \mathbf{x} \\
                                                       \mathbf{y}_{2}\\
                                                     \end{array}
                                                   \right)}{{\mathbf{x}^H}\mathbf{x}}
             \end{split}
\end{equation}
where $\bm\Sigma$ is defined as
\begin{equation}\label{equ:AppendixA7}
\begin{split}
 \bm\Sigma&=\left(\begin{array}{cc} \bm\Pi^{-1} & \mathbf{0} \\
                                \mathbf{0} & \mathbf{I}_{(n-\varsigma)}
              \end{array}\right)\mathbf{U}^H{\mathbf{B}}\mathbf{U}
              \left(\begin{array}{cc}
              \bm\Pi^{-1} & \mathbf{0} \\
              \mathbf{0} & \mathbf{I}_{(n-\varsigma)}\end{array}\right)^{H}\\
              &=\left(\begin{array}{cc} \bm\Pi^{-1} & \mathbf{0} \\
                                \mathbf{0} & \mathbf{I}_{(n-\varsigma)}
              \end{array}\right) \left(\begin{array}{cc}
                                         \widetilde{\mathbf{B}}_{0} & \widetilde{\mathbf{B}}_{2} \\
                                         \widetilde{\mathbf{B}}_{2}^{H} & \widetilde{\mathbf{B}}_{1}
                                       \end{array}
              \right)
              \left(\begin{array}{cc}
              \bm\Pi^{-1} & \mathbf{0} \\
              \mathbf{0} & \mathbf{I}_{(n-\varsigma)}\end{array}\right)^{H}\\
              &= \left(\begin{array}{cc}
                                         \bm\Pi^{-1}\widetilde{\mathbf{B}}_{0}(\bm\Pi^{-1})^{H} & \bm\Pi^{-1}\widetilde{\mathbf{B}}_{2} \\
                                         (\bm\Pi^{-1}\widetilde{\mathbf{B}}_{2})^{H} & \widetilde{\mathbf{B}}_{1}
                                       \end{array}
              \right)\triangleq\left(\begin{array}{cc}
                              \bm\Sigma_{0} & \bm\Sigma_{2} \\
                              \bm\Sigma_{2}^{H} & \bm\Sigma_{1}\end{array}\right).
\end{split}
\end{equation}
\par If $\mathbf{x}$ is a solution to (\ref{equ:AppendixA6}), then so is $\frac{\mathbf{x}}{\|\mathbf{x}\|}$. Thus we can focus on the solutions satisfying  $\mathbf{x}^{H}\mathbf{x}=1$. Then the original problem (\ref{equ:theorem1}) is equivalent to
\begin{equation}\label{equ:AppendixA8}
  \min\limits_{\|\mathbf{x}\|^{2}=1}~~\left(\begin{array}{cc}
                                           \mathbf{x}^{H}~\mathbf{y}_{2}^{H}
                                         \end{array}\right)
                                         \bm\Sigma\left(\begin{array}{c}
                                                        \mathbf{x} \\
                                                       \mathbf{y}_{2}\\
                                                     \end{array}
                                                   \right).
\end{equation}
\par The Lagrange function of (\ref{equ:AppendixA8}) can be written as
\begin{equation}\label{equ:AppendixA9}
\begin{split}
 L(\mathbf{x},\mathbf{y}_{2},\lambda)&=\mathbf{x}^{H}\bm\Sigma_{0}\mathbf{x}+\mathbf{y}_{2}^{H}\bm\Sigma_{2}^{H}\mathbf{x}+\mathbf{x}^{H} \bm\Sigma_{2}\mathbf{y}_{2}\\
 &~~~+\mathbf{y}_{2}^{H}\bm\Sigma_{1}\mathbf{y}_{2}+\lambda(1-\|\mathbf{x}\|^{2}).
 \end{split}
\end{equation}
\par To get the stationary points of $L(\mathbf{x},\mathbf{y}_{2},\lambda)$, we use the theorems in \cite{Brandwood1983},
where for a real function $L$ of the complex variable vectors $\mathbf{\xi}$ and $\mathbf{\xi}^{*}$, the stationary points
 are obtained by solving $\nabla_{\mathbf{\mathbf{\xi}}}L=0$ or $\nabla_{\mathbf{\mathbf{\xi}^{*}}}L=\mathbf{0}$.
 Here we use the gradients of the conjugates of the vectors to obtain the necessary condition for a stationary point as
 \begin{align}
 \nabla_{\mathbf{x}^{*}}
 L(\mathbf{x},\mathbf{y}_{2},\lambda)&=\bm\Sigma_{0}\mathbf{x}+\bm\Sigma_{2}\mathbf{y}_{2}-\lambda\mathbf{x}=\mathbf{0}\label{equ:AppendixA11a}\\
 \nabla_{\mathbf{y}_{2}^{*}} L(\mathbf{x},\mathbf{y}_{2},\lambda)&=\bm\Sigma_{2}^{H}\mathbf{x}+\bm\Sigma_{1}\mathbf{y}_{2}=\mathbf{0}\label{equ:AppendixA11b}\\
 \nabla_{\lambda}L(\mathbf{x},\mathbf{y}_{2},\lambda)&=1-\|\mathbf{x}\|^{2}=0\label{equ:AppendixA11c}
 \end{align}
\par On the other hand, let $\bm\Gamma=\mathbf{A}+\mathbf{B}$, and left and right multiply the equation with $\mathbf{U}^{H}$ and $\mathbf{U}$, respectively, then we have
\begin{equation}\label{equ:AppendixA12}
\begin{split}
\mathbf{U}^{H}\bm\Gamma\mathbf{U}&=\mathbf{U}^{H}\mathbf{A}\mathbf{U}+\mathbf{U}^{H}\mathbf{B}\mathbf{U}\\
&=\left(\begin{array}{cc}
                     \bm\Lambda_{1} & \mathbf{0} \\
                     \mathbf{0} & \mathbf{0 }
                   \end{array}\right)+
    \left(\begin{array}{cc}
                                         \widetilde{\mathbf{B}}_{0} & \widetilde{\mathbf{B}}_{2} \\
                                         \widetilde{\mathbf{B}}_{2}^{H} & \widetilde{\mathbf{B}}_{1}
                                       \end{array}
              \right)\\
              &=\left(\begin{array}{cc}
                                         \bm\Lambda_{1}+\widetilde{\mathbf{B}}_{0} & \widetilde{\mathbf{B}}_{2} \\
                                         \widetilde{\mathbf{B}}_{2}^{H} & \widetilde{\mathbf{B}}_{1}
                                       \end{array}
              \right).
\end{split}
\end{equation}
 \par Since $\bm\Gamma=\mathbf{A}+\mathbf{B}$ is positive definite, then the matrix on the right-hand side (RHS) of (\ref{equ:AppendixA12}) is positive definite, thus $\bm\Sigma_{1}=\widetilde{\mathbf{B}}_{1}$
 should also be positive definite. Therefore based on (\ref{equ:AppendixA11b}) we have
 $\mathbf{y}_{2}=-{\bm\Sigma_{1}^{-1}}{\bm\Sigma_{2}^{H}}\mathbf{x}$. Then combining (\ref{equ:AppendixA11a}) and (\ref{equ:AppendixA11c}), all the stationary points should satisfy
 \begin{equation}\label{euq:AppendixA12a}
 \begin{split}
  \left(\bm\Sigma_{0}-\bm\Sigma_{2}{\bm\Sigma_{1}^{-1}}{\bm\Sigma_{2}^{H}}\right)\mathbf{x}&=\lambda\mathbf{x} \\
   \mathbf{y}_{2}&=-{\bm\Sigma_{1}^{-1}}{\bm\Sigma_{2}^{H}}\mathbf{x}\\
   \|\mathbf{x}\|^{2}&=1
 \end{split}
\end{equation}
\par To obtain the minimal point of (\ref{equ:AppendixA8}), we can find $(\mathbf{x},
 \mathbf{y}_{2},\lambda)$ among the stationary points (i.e., the constraint of (\ref{euq:AppendixA12a})) to make
 the objective function in (\ref{equ:AppendixA8}) the least by
 \begin{equation}\label{euq:AppendixA12b}
 \begin{split}
  \min\limits_{\mathbf{x},\mathbf{y}_{2},\lambda}&~~ \left(\begin{array}{cc}
                                           \mathbf{x}^{H}~\mathbf{y}_{2}^{H}
                                         \end{array}\right)
                                         \bm\Sigma\left(\begin{array}{c}
                                                        \mathbf{x} \\
                                                       \mathbf{y}_{2}\\
                                                     \end{array}
                                                   \right)\\
  \mathrm{s.t.}&~~\left(\bm\Sigma_{0}-\bm\Sigma_{2}{\bm\Sigma_{1}^{-1}}{\bm\Sigma_{2}^{H}}\right)\mathbf{x}=\lambda\mathbf{x} \\
   &~~\mathbf{y}_{2}=-{\bm\Sigma_{1}^{-1}}{\bm\Sigma_{2}^{H}}\mathbf{x}\\
   &~~\|\mathbf{x}\|^{2}=1
   \end{split}
\end{equation}
 \par Left multiplying the first constraint of (\ref{euq:AppendixA12b}) by $\mathbf{x}^{H}$ on both sides and inserting the first two constraints into the objective function, then (\ref{euq:AppendixA12b}) can be simplified by
 \begin{equation}\label{euq:AppendixA12c}
\min\limits_{\|\mathbf{x}\|^{2}=1}~~\mathbf{x}^{H}\left(\bm\Sigma_{0}-{\bm\Sigma_{2}}{\bm\Sigma_{1}^{-1}}{\bm\Sigma_{2}^{H}}\right)\mathbf{x}\\
\end{equation}
\par Based on the Rayleigh-Ritz theorem \cite[Sec. 4.2.2]{Horn1985}, then the minimal solution of (\ref{euq:AppendixA12c}) becomes the eigenvector corresponding to the minimum eigenvalue of $({\bm\Sigma_{0}}- {\bm\Sigma_{2}}{\bm\Sigma_{1}^{-1}}{\bm\Sigma_{2}^{H}})$, where $({\bm\Sigma_{0}}- {\bm\Sigma_{2}}{\bm\Sigma_{1}^{-1}}{\bm\Sigma_{2}^{H}})={\bm\Lambda_{1}^{-\frac{1}{2}}}\left(\widetilde{\mathbf{B}}_{0}-{{\widetilde{\mathbf{B}}}_{2}}{{\widetilde{\mathbf{B}}}_{1}^{-1}} {{\widetilde{\mathbf{B}}}_{2}^{H}}\right){\left(\bm\Lambda _{1}^{-\frac{1}{2}}\right)^H}$ based on (\ref{equ:AppendixA7}). If we assume $\mathbf{x}_{min}$ is the eigenvector, then the minimum value of (\ref{equ:theorem1}) is achieved by
\begin{equation}\label{euq:AppendixA15}
\begin{split}
  \mathbf{r}&=\mathbf{U}\left(\begin{array}{c}
                        \mathbf{y}_{1} \\
                        \mathbf{y}_{2}
                      \end{array}
  \right)=\mathbf{U}\left(\begin{array}{c}
                        (\bm\Pi^{-1})^{H}\mathbf{x}_{min} \\
                       - \widetilde{\mathbf{B}}_{1}^{-1}\widetilde{\mathbf{B}}_{2}^{H}{\left(\bm\Pi^{ - 1} \right)^H}\mathbf{x}_{min}
                      \end{array}
  \right)\\
  &=\mathbf{U}\left( {\begin{array}{*{20}{c}}
\left(\bm\Lambda_{1}^{-\frac{1}{2}}\right)^H\\
{ - {\widetilde{\mathbf{B}}_{1}^{- 1}}\widetilde{\mathbf{B}}_{2}^{H} \left(\bm\Lambda_{1}^{ - \frac{1}{2}}\right)^H}
\end{array}} \right)\mathbf{x}_{min}.
\end{split}
\end{equation}
2){$\varsigma=n$}
\par When $\varsigma=n$, $\mathbf{A}$ is a positive definite matrix and $\bm\Lambda_{1}\in \mathbf{\mathbb{R}}^{n\times n}$.
Using the same process as for the case when $0<\varsigma<n$, we obtain that the minimum value of (\ref{equ:theorem1}) is \cite{Beck2010}
\begin{equation}\label{euq:AppendixA16}
   \mathbf{r}=\left(\mathbf{A}^{-1/2}\right)^{H}\mathbf{x}_{min}
\end{equation}
where $\mathbf{x}_{min}$ is the eigenvector corresponding to the minimal eigenvalue of $\mathbf{A}^{-1/2}\mathbf{B}(\mathbf{A}^{-1/2})^{H}$.
\par Therefore the proof of Theorem 1 is completed. $\hfill\blacksquare$
\subsection{Beampattern formation for desired beamwidths}\label{sec:SubSecBPDB}
\par In the following, we will propose different convex methods to solve the problem of (\ref{equ:optimizationBW}) according to the relationship between the number of transmitters $M$ and transmit waveforms $Q$.
\subsubsection{Cases for $Q=M$} \label{subsec:secA} Since $Q=M$, we can use a Hermitian semi-definite matrix $\mathbf{R}$ ($\in\mathbf{\mathbb{C}}^{M\times M}$) to represent $\mathbf{C}\mathbf{C}^{H}={\sum_{q=1}^{M}{\mathbf{c}_{q}\mathbf{c}^{H}_{q}}}$ without loss of any generality. Then (\ref{equ:BeampatternTranfo}) can be reformulated as
\begin{equation}\label{equ:beampattn1}
\begin{split}
  P(\theta)&=\mathbf{c}^{H}\left( {\mathbf{I}_Q}\otimes {\mathbf{A}(\theta)} \right)\mathbf{c}=\sum_{q=1}^{M}{\mathbf{c}_{q}^{H}{\mathbf{A}(\theta)}\mathbf{c}_{q}}\\
  &=\mathrm{tr}\left\{{\mathbf{A}(\theta)}\sum_{q=1}^{M}{\mathbf{c}_{q}\mathbf{c}_{q}^{H}}\right\}=\mathrm{tr}\left\{{\mathbf{A}(\theta)}\mathbf{R}\right\}
\end{split}
\end{equation}
\par Thus (\ref{equ:optimizationBW}) can be reformulated as
\begin{equation}\label{equ:NonConvex1}
\begin{split}
\min\limits_{\mathbf{R}}&~~\frac{\mathrm{tr}\left\{{\mathbf{A}_{M}^{sl}}\mathbf{R}\right\}}{\mathrm{tr}\left\{{\mathbf{A}_{M}^{ml}}\mathbf{R}\right\}}\\
\mathrm{s.t.}&~~\frac{1}{2}\leq\frac{\mathrm{tr}\left\{{\mathbf{A}(\theta)}\mathbf{R}\right\}}{\mathrm{tr}\left\{{\mathbf{A}(\theta_{0})}\mathbf{R}\right\}}\leq1,~~\forall\theta\in\Theta_{ml}\\
             &~~\mathbf{R}\succeq 0
\end{split}
\end{equation}
where $\mathbf{A}_{M}^{ml}\triangleq{\int_{\Theta_{ml}}{\mathbf{A}(\theta)}{d\theta}}$, $\mathbf{A}_{M}^{sl} \triangleq{\int_{\Theta_{sl}}{\mathbf{A}(\theta)}{d\theta}}$, { $\mathbf{A}_{ml}=\mathbf{I}_{Q}\otimes\mathbf{A}_{M}^{ml}$ and $\mathbf{A}_{sl}=\mathbf{I}_{Q}\otimes\mathbf{A}_{M}^{sl}$. Also $\succeq$ in (\ref{equ:NonConvex1}) denotes that $\mathbf{R}$ should be a semi-definite matrix.}
\par The objective function of (\ref{equ:NonConvex1}) is a ratio of two linear functions of $\mathbf{R}$  Thus if $\mathbf{R}$ is a solution to (\ref{equ:NonConvex1}), so is $\frac{\mathbf{R}}{\mathrm{tr}\{A_{M}^{ml}\mathbf{R}\}}$. Therefore we can focus on the solution {that} satisfies $\mathrm{tr}\{A_{M}^{ml}\mathbf{R}\}=1$. Then (\ref{equ:NonConvex1}) is equivalent to
\begin{equation}\label{equ:Convex1}
\begin{split}
\min\limits_{\mathbf{R}}&~~\mathrm{tr}\left\{{\mathbf{A}_{M}^{sl}}\mathbf{R}\right\}\\
\mathrm{s.t.}&~~\mathrm{tr}\left\{{\mathbf{A}_{M}^{ml}}\mathbf{R}\right\}=1\\
             &~~\frac{1}{2}\leq\frac{\mathrm{tr}\left\{{\mathbf{A}(\theta)}\mathbf{R}\right\}}{\mathrm{tr}\left\{{\mathbf{A}(\theta_{0})}\mathbf{R}\right\}}\leq1,~~\forall\theta\in\Theta_{ml}\\
             &~~\mathbf{R}\succeq 0
\end{split}
\end{equation}
\par Obviously, (\ref{equ:Convex1}) is a semi-definite programming (SDP)\cite{Vandenberghe1996} problem for the variable $\mathbf{R}$. Thus we can use the convex optimization toolbox CVX \cite{Grant2008,CVXResearch2012} directly to get the optimal solution.
\subsubsection{Cases for $Q\neq M$}
Since $\mathbf{x}^{H}\mathbf{T}\mathbf{x}=\mathrm{tr}\{\mathbf{T}\mathbf{x}\mathbf{x}^{H}\}$ for $\forall \mathbf{x} $ and $\mathbf{T}$, then  (\ref{equ:optimizationBW}) {is} equivalent to
 \begin{equation}\label{equ:optBW1}
\begin{split}
\min\limits_{\mathbf{c},\mathbf{X}}&~~\mathrm{tr}\{{\mathbf{A}_{sl}}\mathbf{X}\}\\
\mathrm{s.t.}&~~\mathrm{tr}\{{\mathbf{A}_{ml}}\mathbf{X}\}=1\\ &~~\frac{1}{2}\leq\frac{\mathrm{tr}\{\widetilde{\mathbf{A}}(\theta)\mathbf{X}\}}{\mathrm{tr}\{\widetilde{\mathbf{A}}(\theta_{0})\mathbf{X}\}}\leq1,~~\forall\theta\in\Theta_{ml}\\
&~~\mathbf{X}=\mathbf{c}\mathbf{c}^{H}
\end{split}
\end{equation}
\par Observing (\ref{equ:optBW1}), we find that it is not a convex optimization due to the last nonconvex constraint $\mathbf{X}=\mathbf{c}\mathbf{c}^{H}$ ($\in\mathbf{\mathbb{C}}^{MQ\times MQ}$). Therefore, {to solve} (\ref{equ:optBW1}), we consider using semidefinite relaxation (SDR) \cite{Ma2002,2003}. Relaxing the constraint $\mathbf{X}=\mathbf{c}\mathbf{c}^{H}$ {to} a convex positive semidefinite constraint $\mathbf{X}-\mathbf{c}\mathbf{c}^{H}\succeq 0$, we can obtain a lower bound on the optimal value of (\ref{equ:optBW1}) by solving the following convex problem
\begin{equation}\label{equ:Convex3}
\begin{split}
\min\limits_{\mathbf{X},\mathbf{c}}&~~\mathrm{tr}\left\{{\mathbf{A}_{sl}}\mathbf{X}\right\}\\
\mathrm{s.t.}&~~\mathrm{tr}\left\{{\mathbf{A}_{ml}}\mathbf{X}\right\}=1\\
             &~~\frac{1}{2}\leq\frac{\mathrm{tr}\left\{\widetilde{\mathbf{A}}(\theta)\mathbf{X}\right\}}{\mathrm{tr}\left\{{\widetilde{\mathbf{A}}(\theta_{0})}\mathbf{X}\right\}}\leq1,~~\forall\theta\in\Theta_{ml}\\
             &~~\mathbf{X}-{\mathbf{c}\mathbf{c}^{H}}\succeq 0
\end{split}
\end{equation}
\par Hence problem (\ref{equ:Convex3}) is the SDR of (\ref{equ:optimizationBW}) and can be solved with CVX. Suppose the optimal solution of (\ref{equ:Convex3}) is $(\mathbf{X}_{\star},\mathbf{c}_{\star})$. This may not be the solution of (\ref{equ:optimizationBW}).
To get an accurate approximate solution for (\ref{equ:optimizationBW}) based on $(\mathbf{X}_{\star},\mathbf{c}_{\star})$,  Gaussian randomization \cite{2003} can be used. Generate a sufficient number of samples by assuming $\mathbf{c}$ is a Gaussian variable with $\mathbf{c}\sim \mathcal{CN}(\mathbf{c}_{\star}, \mathbf{X}_{\star}-\mathbf{c}_{\star}\mathbf{c}_{\star}^{H})$, {then choose} the samples satisfying the constraint of (\ref{equ:optimizationBW}) to find the best feasible point among them to make the objective function the smallest.
\section{Numerical Results} \label{sec:NumSim}
In this section, some theoretical and numerical analyses are performed to show how the variables $Q$ (number of waveforms) and $M$ (number of transmitters) impact the designed beampatterns. Next, some comparison simulations are conducted to verify the performance our methods.
\subsection{Beampattern analyses with respect to the number of waveforms $Q$ and transmitters $M$}\label{Sec:subsecNR_1}
\subsubsection{Analysis on the impact of different $Q$ on the designed beampatterns}
First we give a Lemma which will be useful to our analyses.
 \par \textbf{Lemma 2:} Given Hermitian non-negative definite matrices $\mathbf{A},\mathbf{B}\in \mathbf{\mathbb{C}}^{n\times n}$, if there exists $k_{min}(\geq 0)$ such that $\min\limits_{\mathbf{r}\in \mathbf{\mathbb{C}}^{n\times1}}\frac{{{\mathbf{r}^H}{\mathbf{B}}\mathbf{r}}}{{{\mathbf{r}^H}{\mathbf{A}}\mathbf{r}}}=k_{min}(\geq0)$, then
  $k_{min}=\min\limits_{\mathbf{r}\in \mathbf{\mathbb{C}}^{n\times1}}\frac{{{\mathbf{r}^H}{\mathbf{B}}\mathbf{r}}}{{{\mathbf{r}^H}{\mathbf{A}}\mathbf{r}}}
     =\min\limits_{\mathbf{X}\succeq 0} \frac{\mathrm{tr}\{\mathbf{B}\mathbf{X}\}}{\mathrm{tr}\{\mathbf{A}\mathbf{X}\}}$, where $\mathbf{X}(\in\mathbf{\mathbb{C}}^{n\times n})$ is a positive semidefinite Hermitian matrix.
     \par
\begin{figure}[ht]
  \subfigure[]
  {\begin{minipage}{1\textwidth}
  \centering
  \includegraphics[scale=0.64]{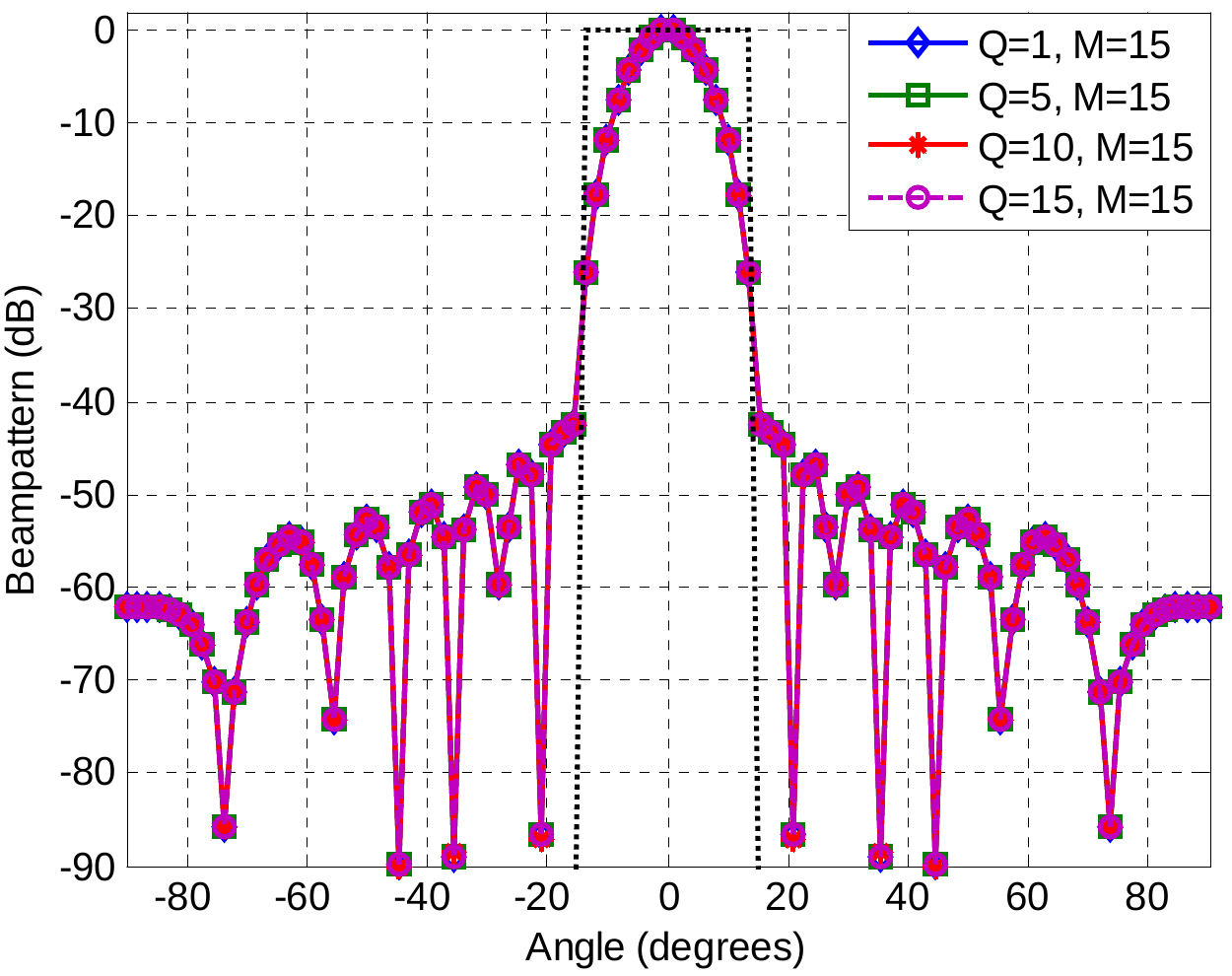}\label{subfig:2a}
  \end{minipage}}\\
  \subfigure[]
  {\begin{minipage}{1\textwidth}
  \centering
  \includegraphics[scale=0.64]{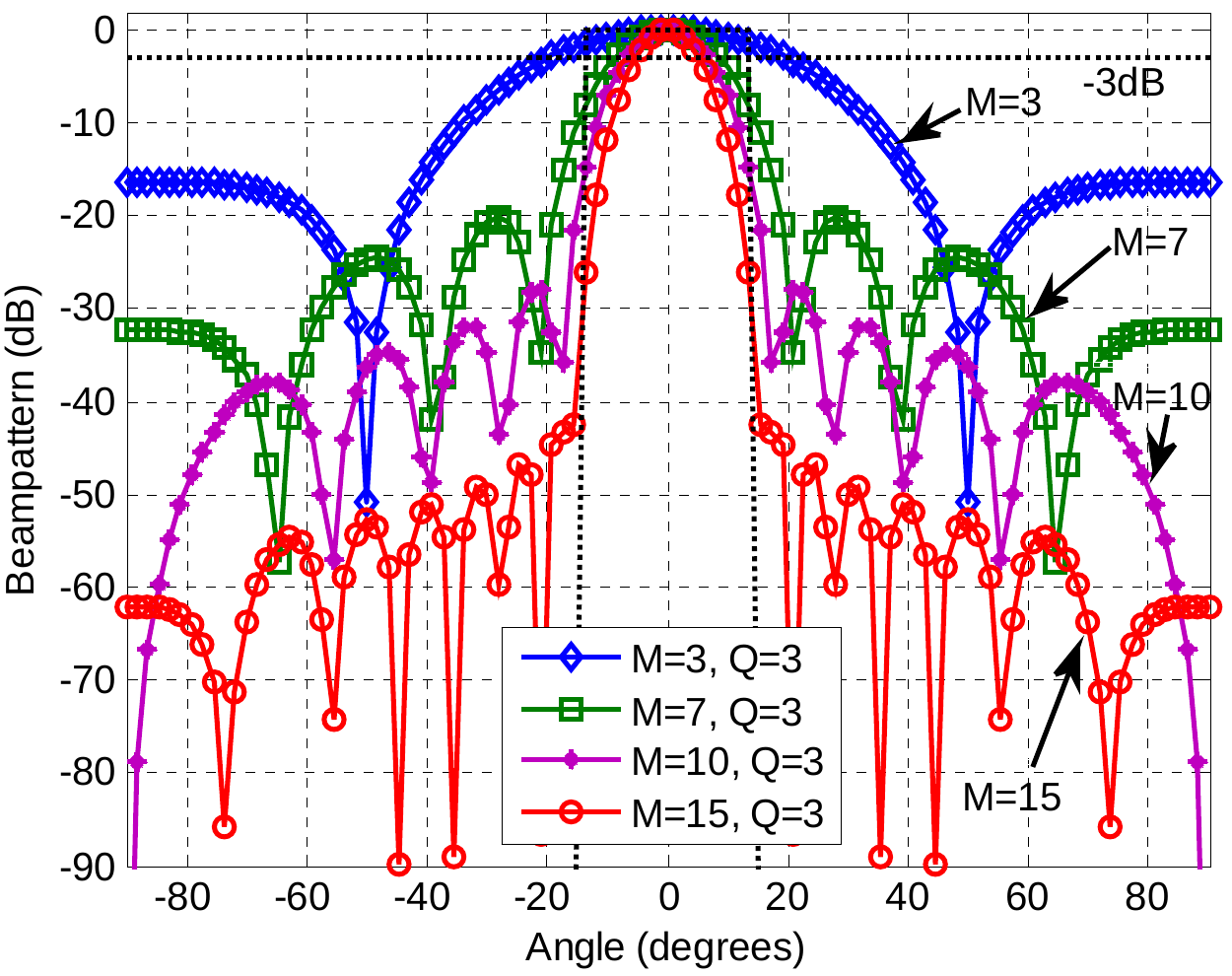}\label{subfig:2b}
  \end{minipage}}\\
  \caption{Normalized beampatterns designed by the analytical method for a desired $30^{\circ}$ width
  of the main lobe, (a) vs. different values of $Q$; (b) vs. different values of $M$.}\label{fig:2}
\end{figure}

 See the proof in Appendix \ref{App:ProofLemma2}.
Given a non-zero vector $\mathbf{c}_{+}=[\mathbf{c}_{1}^{T}~\mathbf{c}_{2}^{T}]^{T}\in \mathbf{\mathbb{C}}^{M(Q+1)\times1}$, where $\mathbf{c}_{1}\in\mathbf{\mathbb{C}}^{MQ\times1}$ and $\mathbf{c}_{2}\in\mathbf{\mathbb{C}}^{M\times1}$, we have
\begin{equation}\label{equ:SimuR1}
\begin{split}
&\mathbf{c}_{+}^{H}\left(\mathbf{I}_{Q+1}\otimes \mathbf{A}_{M}^{sl}\right)\mathbf{c}_{+}\\
&=\mathbf{c}_{1}^{H}\left(\mathbf{I}_{Q}\otimes \mathbf{A}_{M}^{sl}\right)\mathbf{c}_{1}+\mathbf{c}_{2}^{H}\mathbf{A}_{M}^{sl}\mathbf{c}_{2}\\
&=\mathbf{c}_{1}^{H}\left(\mathbf{I}_{Q}\otimes \mathbf{A}_{M}^{sl}\right)\mathbf{c}_{1}+\widetilde{\mathbf{c}}_{2}^{H}\left(\mathbf{I}_{Q}\otimes\mathbf{A}_{M}^{sl}\right)\widetilde{\mathbf{c}}_{2}\\
&=\mathrm{tr}\left\{\left(\mathbf{I}_{Q}\otimes\mathbf{A}_{M}^{sl}\right)\left(\mathbf{c}_{1}\mathbf{c}_{1}^{H}+\widetilde{\mathbf{c}}_{2}\widetilde{\mathbf{c}}_{2}^{H}\right)\right\}
\end{split}
\end{equation}
where $\widetilde{\mathbf{c}}_{2}\in\mathbf{\mathbb{C}}^{MQ\times1}$ and $\widetilde{\mathbf{c}}_{2}\triangleq[\mathbf{c}_{2}^{T}~\mathbf{0}_{M(Q-1)}^{T}]^{T}$. Based on (\ref{equ:SimuR1}), we have
\begin{equation}\label{equ:SimuR2}
\begin{split}
   &\min\limits_{\mathbf{c}_{+}\in\mathbf{\mathbb{C}}^{M(Q+1)\times1}}\frac{\mathbf{c}_{+}^{H}\left(\mathbf{I}_{Q+1}\otimes \mathbf{A}_{M}^{sl}\right)\mathbf{c}_{+}}{\mathbf{c}_{+}^{H}\left(\mathbf{I}_{Q+1}\otimes \mathbf{A}_{M}^{ml}\right)\mathbf{c}_{+}}\\
   &=\min\limits_{\mathbf{c}_{+}\in\mathbf{\mathbb{C}}^{M(Q+1)\times1}}
\frac{\mathrm{tr}\left\{\left(\mathbf{I}_{Q}\otimes\mathbf{A}_{M}^{sl}\right)\left(\mathbf{c}_{1}\mathbf{c}_{1}^{H}+\widetilde{\mathbf{c}}_{2}\widetilde{\mathbf{c}}_{2}^{H}\right)\right\}}
{\mathrm{tr}\left\{\left(\mathbf{I}_{Q}\otimes\mathbf{A}_{M}^{ml}\right)\left(\mathbf{c}_{1}\mathbf{c}_{1}^{H}+\widetilde{\mathbf{c}}_{2}\widetilde{\mathbf{c}}_{2}^{H}\right)\right\}}\\
&\geq \min\limits_{\mathbf{X}_{0}\succeq 0}\frac{\mathrm{tr}\left\{\left(\mathbf{I}_{Q}\otimes\mathbf{A}_{M}^{sl}\right)\mathbf{X}_{0}\right\}}
{\mathrm{tr}\left\{\left(\mathbf{I}_{Q}\otimes\mathbf{A}_{M}^{ml}\right)\mathbf{X}_{0}\right\}}
\end{split}
\end{equation}
where $\mathbf{X}_{0}\in \mathbf{\mathbb{C}}^{MQ\times MQ}$ and the inequality holds because the problem on the RHS is a SDR of the problem on the LHS. On the other hand, we have $\min\limits_{\mathbf{c}_{+}\in \mathbf{\mathbb{C}}^{M(Q+1)\times1}}\frac{\mathbf{c}_{+}^{H}\left(\mathbf{I}_{Q+1}\otimes \mathbf{A}_{M}^{sl}\right)\mathbf{c}_{+}}{\mathbf{c}_{+}^{H}\left(\mathbf{I}_{Q+1}\otimes \mathbf{A}_{M}^{ml}\right)\mathbf{c}_{+}}\leq \min\limits_{\mathbf{c}\in \mathbf{\mathbb{C}}^{MQ\times1}} \frac{\mathbf{c}^{H}\left(\mathbf{I}_{Q}\otimes \mathbf{A}_{M}^{sl}\right)\mathbf{c}}{\mathbf{c}^{H}\left(\mathbf{I}_{Q}\otimes \mathbf{A}_{M}^{ml}\right)\mathbf{c}}$ since setting
the extra components in $\mathbf{c}_{+}$ to zero gives the same answer as the problem on the RHS. However, no matter what $Q$ and $M$ are, $\min\limits_{\mathbf{\mathbf{c}}}\frac{\mathbf{c}^{H}(\mathbf{I}_{Q}\otimes\mathbf{A}_{M}^{sl})\mathbf{c}}{\mathbf{c}^{H}(\mathbf{I}_{Q}\otimes\mathbf{A}_{M}^{ml})\mathbf{c}}
=\min\limits_{\mathbf{X}_{0}\succeq 0}\frac{\mathrm{tr}\{(\mathbf{I}_{Q}\otimes\mathbf{A}_{M}^{sl})\mathbf{X}_{0}\}}{\mathrm{tr}\{(\mathbf{I}_{Q}\otimes\mathbf{A}_{M}^{ml})\mathbf{X}_{0}\}}$ will always hold based on Lemma 2.
Thus $\min\limits_{\mathbf{c}_{+}\in\mathbf{\mathbb{C}}^{M(Q+1)\times1}}\frac{\mathbf{c}_{+}^{H}\left(\mathbf{I}_{Q+1}\otimes \mathbf{A}_{M}^{sl}\right)\mathbf{c}_{+}}{\mathbf{c}_{+}^{H}\left(\mathbf{I}_{Q+1}\otimes \mathbf{A}_{M}^{ml}\right)\mathbf{c}_{+}}=
\min\limits_{\mathbf{c}\in\mathbf{\mathbb{C}}^{MQ\times1}}\frac{\mathbf{c}^{H}\left(\mathbf{I}_{Q}\otimes \mathbf{A}_{M}^{sl}\right)\mathbf{c}}{\mathbf{c}^{H}\left(\mathbf{I}_{Q}\otimes \mathbf{A}_{M}^{ml}\right)\mathbf{c}}$. Then we can conclude that the values of $Q$ will not have impact on beampatterns obtained by the analytical method proposed in Section \ref{sec:SubSecBPDM}. Fig. \ref{subfig:2a} has verified our conclusion since all the cases with different $Q$ have the same shaped beampattern. Now we consider the convex method proposed in Section \ref{sec:SubSecBPDB}. To begin our analyses, a lemma is provided.
\par \textbf{Lemma 3:} Given a real-valued function $f(Q)=\min\limits_{\mathbf{X}\succeq 0}{\mathrm{tr}\left\{ {\left( {{\mathbf{I}_Q} \otimes \mathbf{B}} \right)\mathbf{X}} \right\}}$ subject to ${\mathrm{tr}\left\{ {\left( {{\mathbf{I}_Q} \otimes \mathbf{A}_{i}} \right)\mathbf{X}} \right\}}\leq b_{i}~(\in \mathbf{\mathbb{R}})$, $i=1,...,K$, where $\mathbf{A}_{i}$, $\mathbf{B}~(\in \mathbf{\mathbb{C}}^{M\times M})$ and
 $\mathbf{X}~(\in\mathbf{\mathbb{C}}^{MQ\times MQ})$ are all Hermitian matrices\footnote{Here we consider bounded set, since the minimization is a SDP with bounded objective, so the minimal value, i.e., the value of $f$, for each $Q$ should always exist.}, then $f(Q+1)=f(Q)$  holds for $\forall Q\in \mathbf{\mathbb{Z}}^{+}$.
 \par See the proof in Appendix \ref{App:ProofLemma3}. It's not hard for us to prove that the SDRs of the minimization (\ref{equ:optimizationBW}) for a case with $Q+1$ and a case with $Q$ can be expressed in the form of $f(Q+1)$ and $f(Q)$, respectively, as defined in Lemma 3. Therefore, the SDRs for a case with $Q+1$ and a case with $Q$ have the same minimal value based on Lemma 3. Since we find the final solutions for different Q based on different SDRs but with the same minimal value, we can conclude that the beampatterns designed by the convex method for different Q will generally be the same except for minor differences that may exist because we find the best solution for each case among a Gaussian randomized sample. Fig. \ref{subfig:3a} shows the numerical results with different values of $Q$. We can see that there are only minor differences between the different beampatterns, which not only shows the correctness of our conclusion but also shows the good precision of Gaussian randomization.
\subsubsection {Analysis on the impacts of different $M$ on the designed beampatterns} Considering a vector
$\mathbf{c} \in\mathbf{\mathbb{C}}^{M\times1}$, then $\mathbf{c}^{H}\mathbf{A}_{M}^{sl}\mathbf{c}=[\mathbf{c}^{H}~0]
 \mathbf{A}_{M+1}^{sl}[\mathbf{c}^{H}~0]^{H}$ holds. Thus given $\mathbf{c}_{\ast}=[\mathbf{c}_{1}^{T}\cdots \mathbf{c}_{Q}^{T}]^{T}
 \in\mathbf{\mathbb{C}}^{MQ\times1}$, we have
\begin{equation}\label{equ:SimuR2}
\begin{split}
  \mathbf{c}_{\ast}^{H}\left(\mathbf{I}_{Q}\otimes \mathbf{A}_{M}^{sl}\right)\mathbf{c}_{\ast}&={\sum_{q=1}^{Q}{\mathbf{c}_{q}^{H} \mathbf{A}_{M}^{sl} \mathbf{c}_{q}}}\\
  &={\sum_{q=1}^{Q}{[\mathbf{c}_{q}^{H}~0] \mathbf{A}_{M+1}^{sl} [\mathbf{c}_{q}^{H}~0]^{H}}}\\
  &={\widetilde{\mathbf{c}}_{\ast}^{H}\left(\mathbf{I}_{Q}\otimes \mathbf{A}_{M+1}^{sl}\right)\widetilde{\mathbf{c}}_{\ast}}
\end{split}
\end{equation}
where $\widetilde{\mathbf{c}}_{\ast}\in\mathbf{\mathbb{C}}^{(M+1)Q\times1}$ is an extension of $\mathbf{c}_{\ast}$ by
padding a zero at the end of each $M$ elements. Therefore, the beampattern we obtain at $M$ by (\ref{equ:optimization})
 can be viewed as a special case of $M+1$ with $M$ zero-components. Thus when we increase the number of transmitters, we obtain a beampattern with a lower sidelobe level because the case of $M+1$ will provide a lower optimal value than the case of $M$ which means a lower sidelobe level. Fig. \ref{subfig:2b} shows the simulation results of the beampatterns for different values of $M$. From the figure, it can be seen that the sidelobe level of the beampattern decreases when $M$ increases, as the analysis indicated. When $M$ is large enough, the mainlobe width of the beampattern is steady at the desired width of $30^{\circ}$. However, when $M$ is not, though increasing $M$ can reduce sidelobe levels, it will result in a reduction in the
 3dB beamwidth. Thus when using the analytical method proposed in Section \ref{sec:SubSecBPDM}, one should choose $M$ carefully in beampattern formation if it is desirable to obtain a good trade-off between the 3dB beamwidth and the sidelobe level.
 \begin{figure}[ht]
  \subfigure[]
  {\begin{minipage}{1\textwidth}
  \centering
  \includegraphics[scale=0.64]{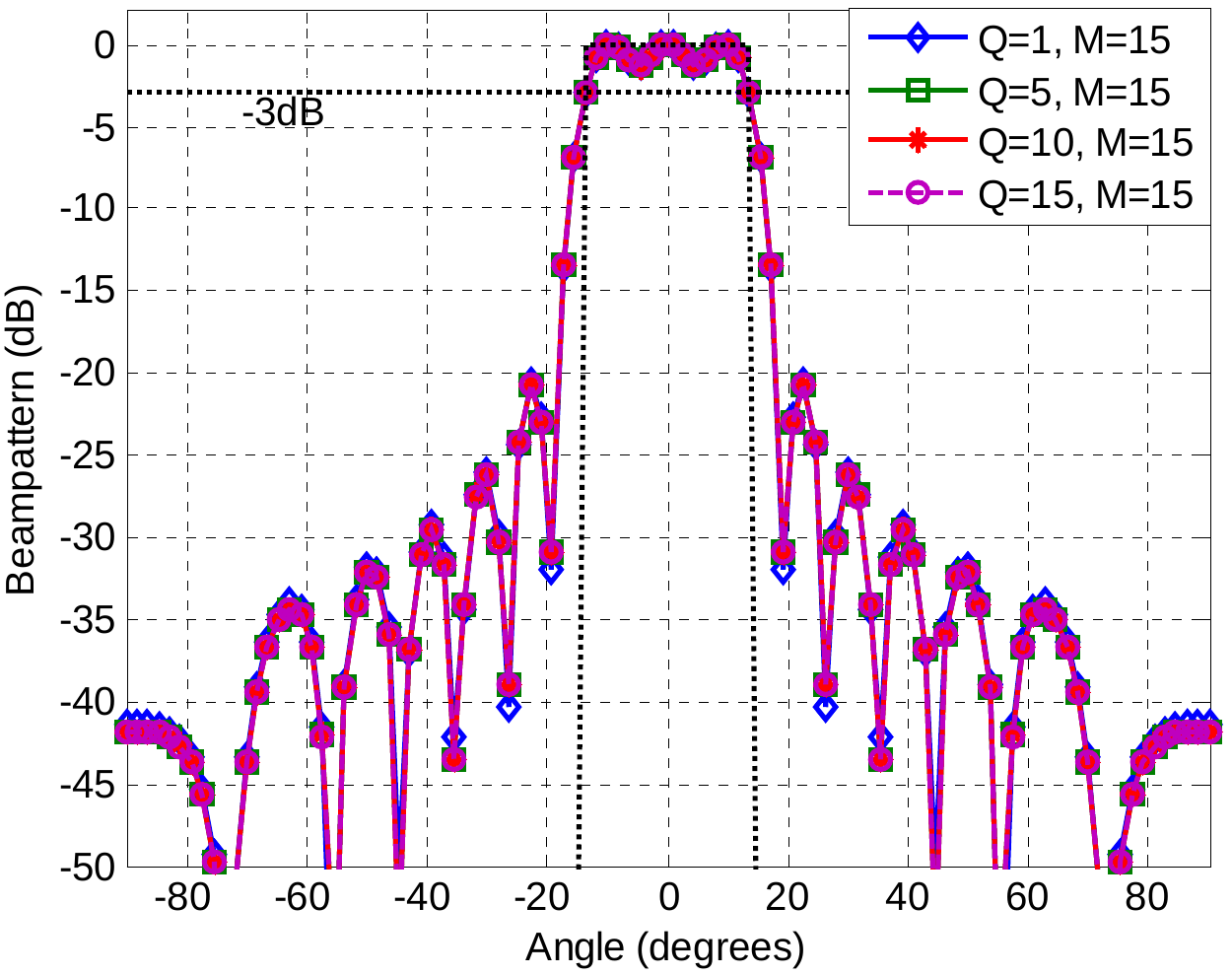}\label{subfig:3a}
  \end{minipage}}\\
  \subfigure[]
  {\begin{minipage}{1\textwidth}
  \centering
  \includegraphics[scale=0.64]{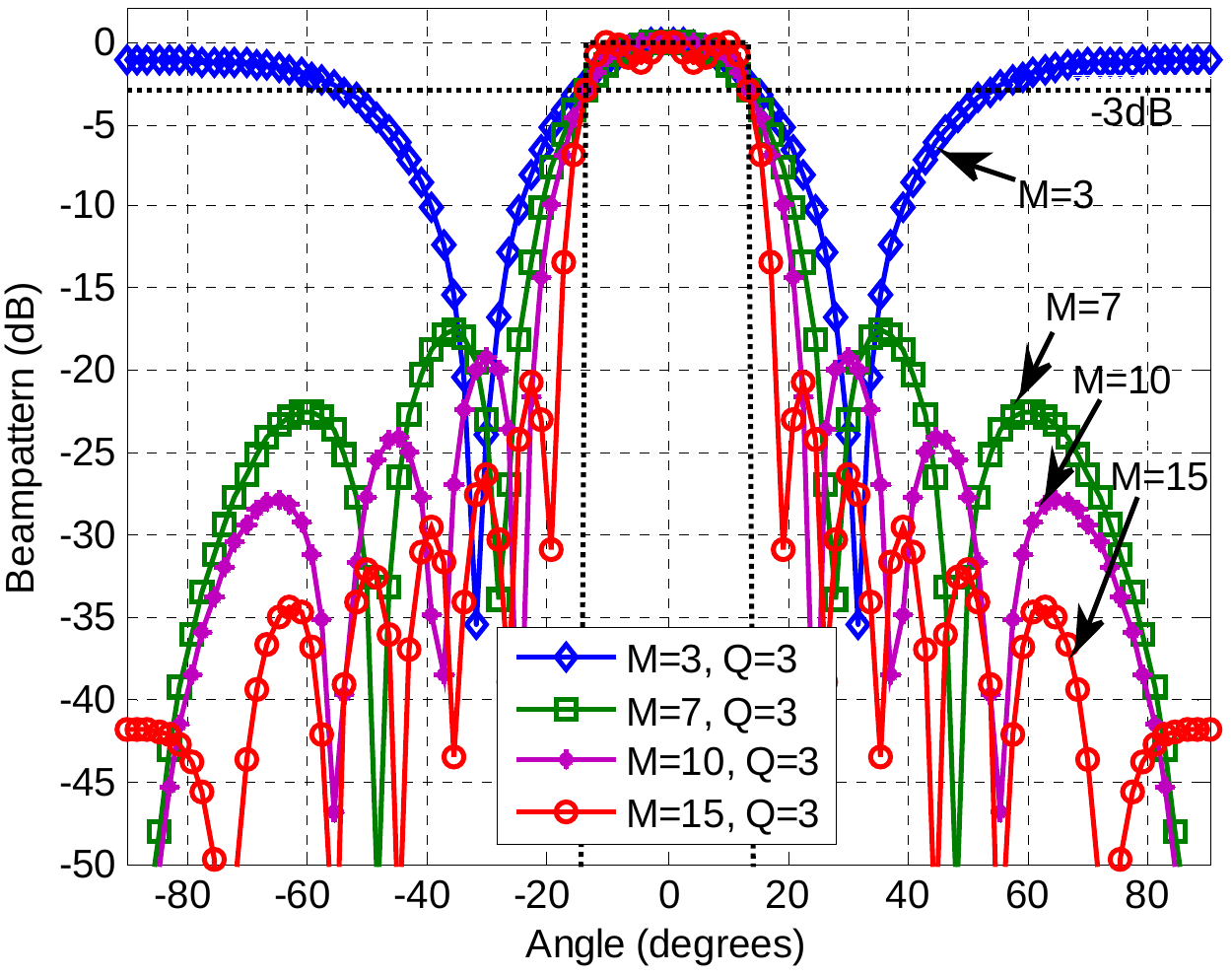}\label{subfig:3b}
  \end{minipage}}\\
  \caption{Normalized beampatterns designed by the convex method with a desired $30^{\circ}$ beamwidth,
  (a) vs. different values of $Q$; (b) vs. different values of $M$.}\label{fig:3}
\end{figure}
Similarly, based on (\ref{equ:SimuR2}) we also
 can convert the case of $M$ for (\ref{equ:optimizationBW}) into a special case of $M+1$ though (\ref{equ:optimizationBW}) has an addtional constraint comparing with (\ref{equ:optimization}),
 which again means that increasing $M$ will lower the sidelobe level. Fig. \ref{subfig:3b}
 has verified our conclusion.
 \begin{figure}[ht!]
  \subfigure[]
  {\begin{minipage}{1\textwidth}
  \centering
  \includegraphics[scale=0.64]{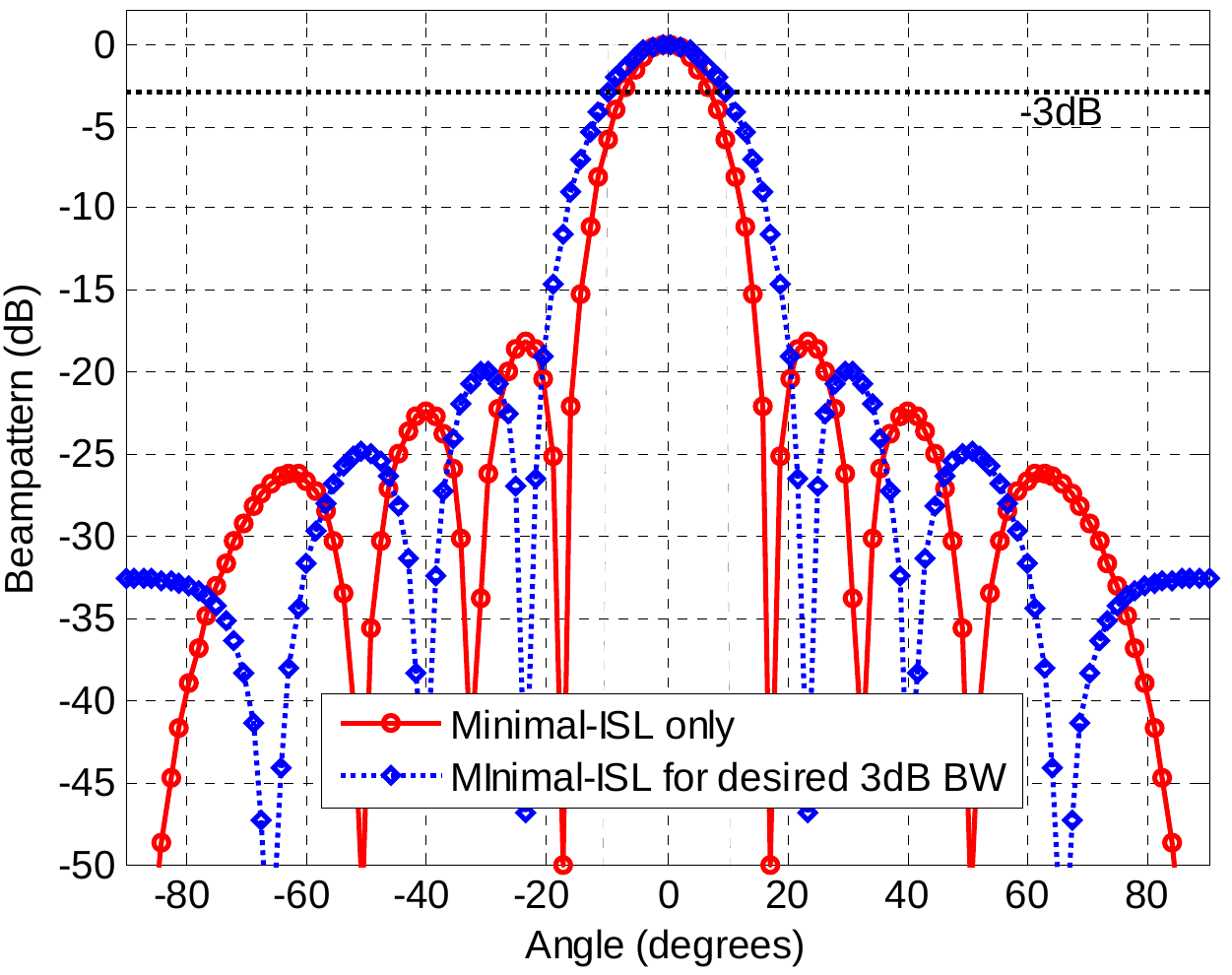}\label{subfig:4a}
  \end{minipage}}\\
  \subfigure[]
  {\begin{minipage}{1\textwidth}
  \centering
  \includegraphics[scale=0.64]{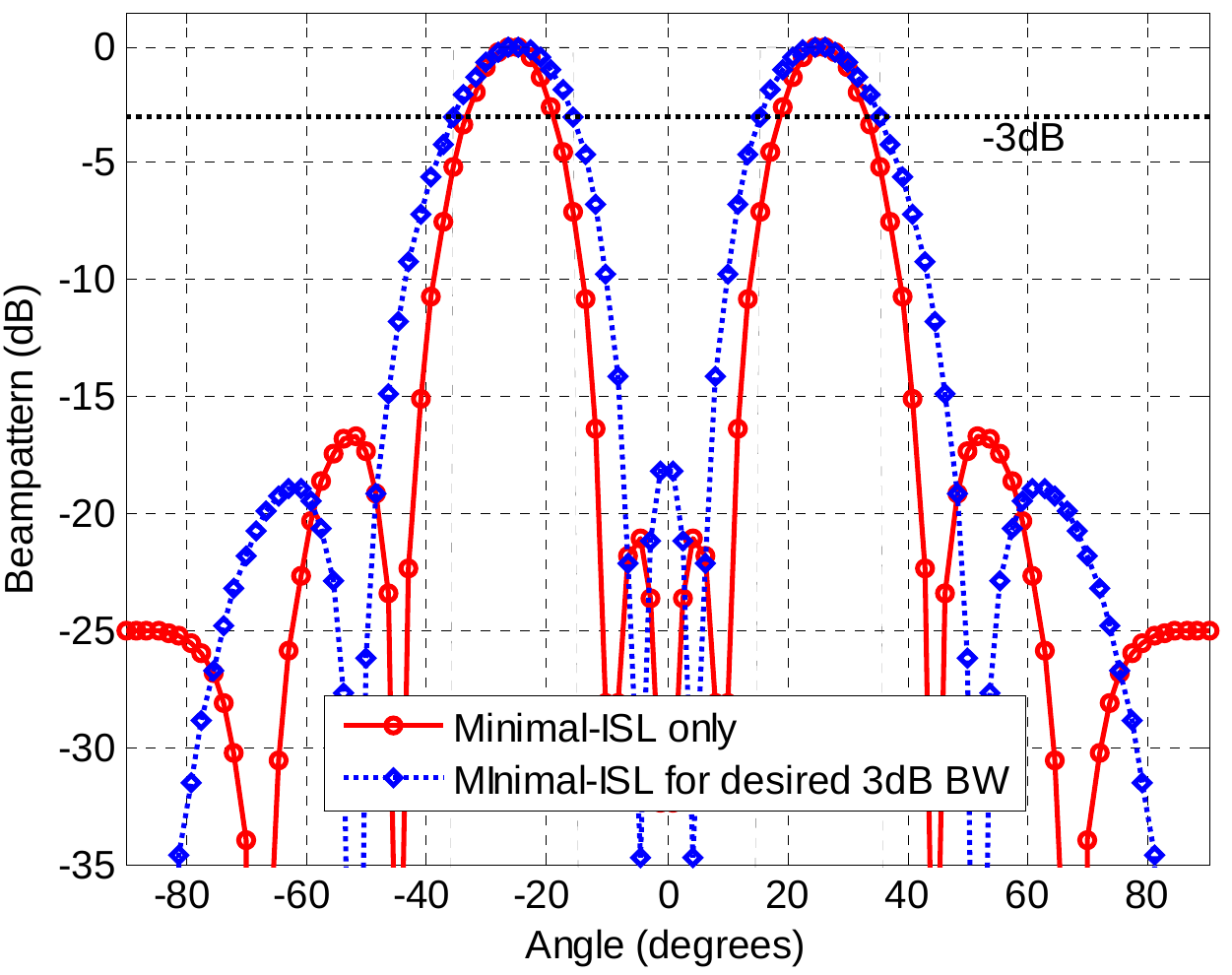}\label{subfig:4b}
  \end{minipage}}\\
   {\begin{minipage}{1\textwidth}
  \centering
  \includegraphics[scale=0.64]{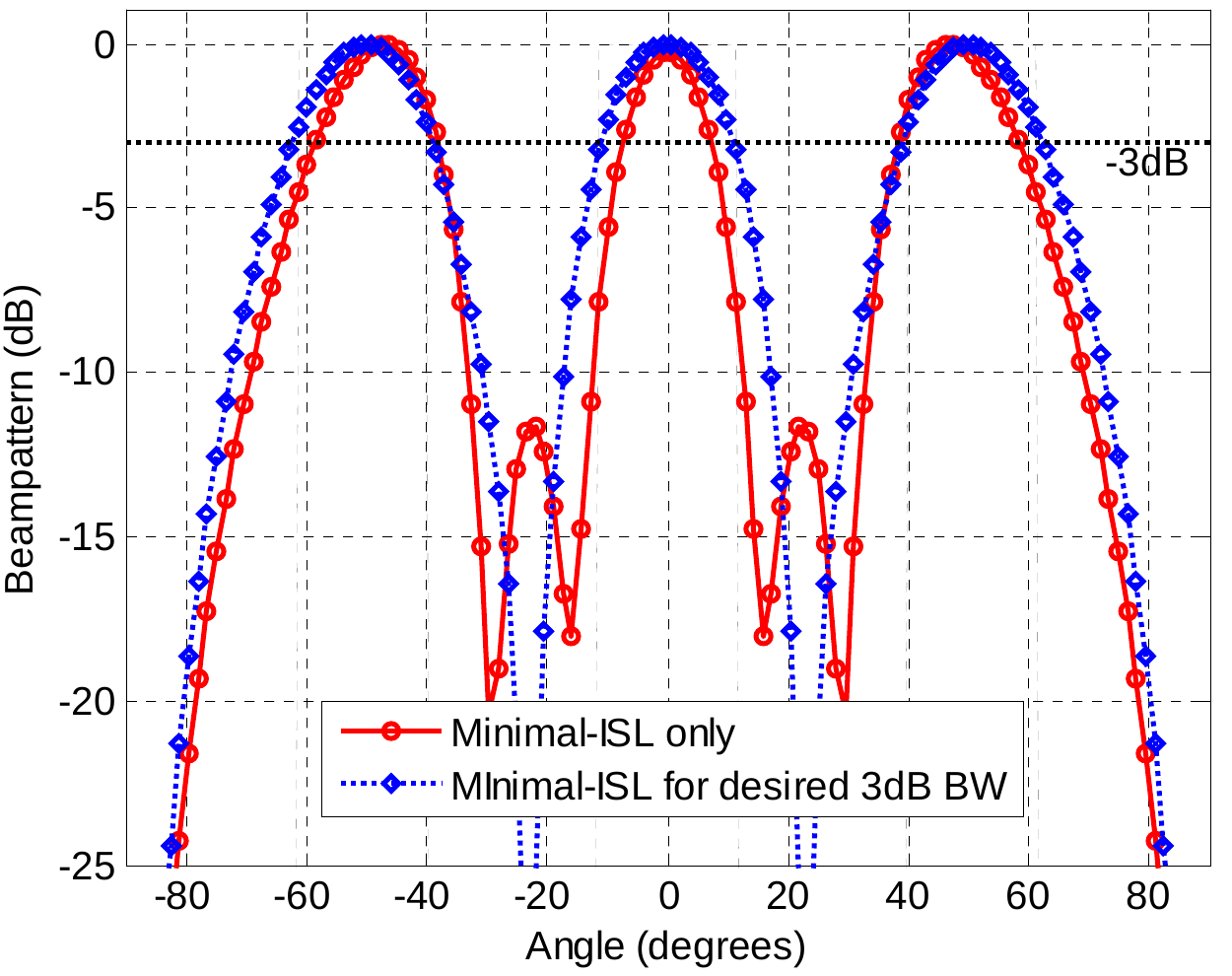}\label{subfig:4c}
  \end{minipage}}\\
  \caption{Beampattern comparisons between the two proposed methods. (a) One-mainlobe case, (b) Two-mainlobe case, (c) Three-mainlobe case.}\label{fig:4_BPCompar1}
\end{figure}
\subsection{Comparisons among the different design methods}
\subsubsection{Comparisons between the two proposed methods} For $M=8$ and $Q=3$, we consider three types of desired beampatterns: the one-mainlobe case, the two-mainlobe case and the three-mainlobe\footnote{Multi-beam parallel design for complicated multi-mainlobe cases can be employed when using the minimal-ISL only criterion.} case. In each case the desired width of each mainlobe region is $22^{\circ}$ while the space between mainlobes is $50^{\circ}$ for the two multiple mainlobe cases. Fig. \ref{fig:4_BPCompar1} shows the numerical results of the beampatterns designed using the two proposed methods. From the three subfigures, it is seen that both of the methods can provide useful beampatterns. Note that all the beampatterns designed by adding the constraint on the 3dB beamwidth achieve a 3dB beamwidth close to that desired as expected.  The mainlobes obtained using the method without the constraint are slightly different as might be expected.
\begin{figure}[ht!]
  \subfigure[]
  {\begin{minipage}{1\textwidth}
  \centering
  \includegraphics[scale=0.64]{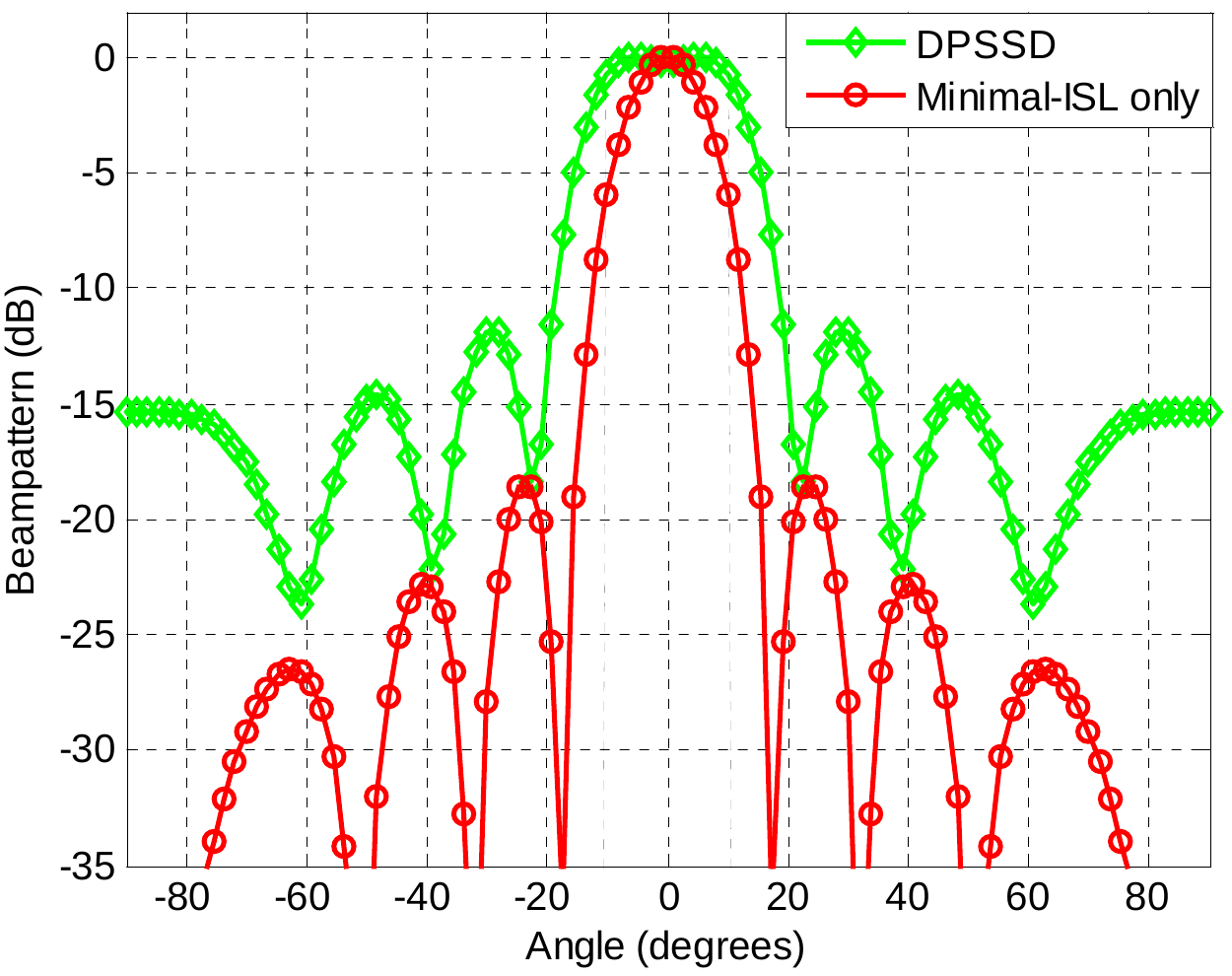}\label{subfig:5a}
  \end{minipage}}\\
  \subfigure[]
  {\begin{minipage}{1\textwidth}
  \centering
  \includegraphics[scale=0.64]{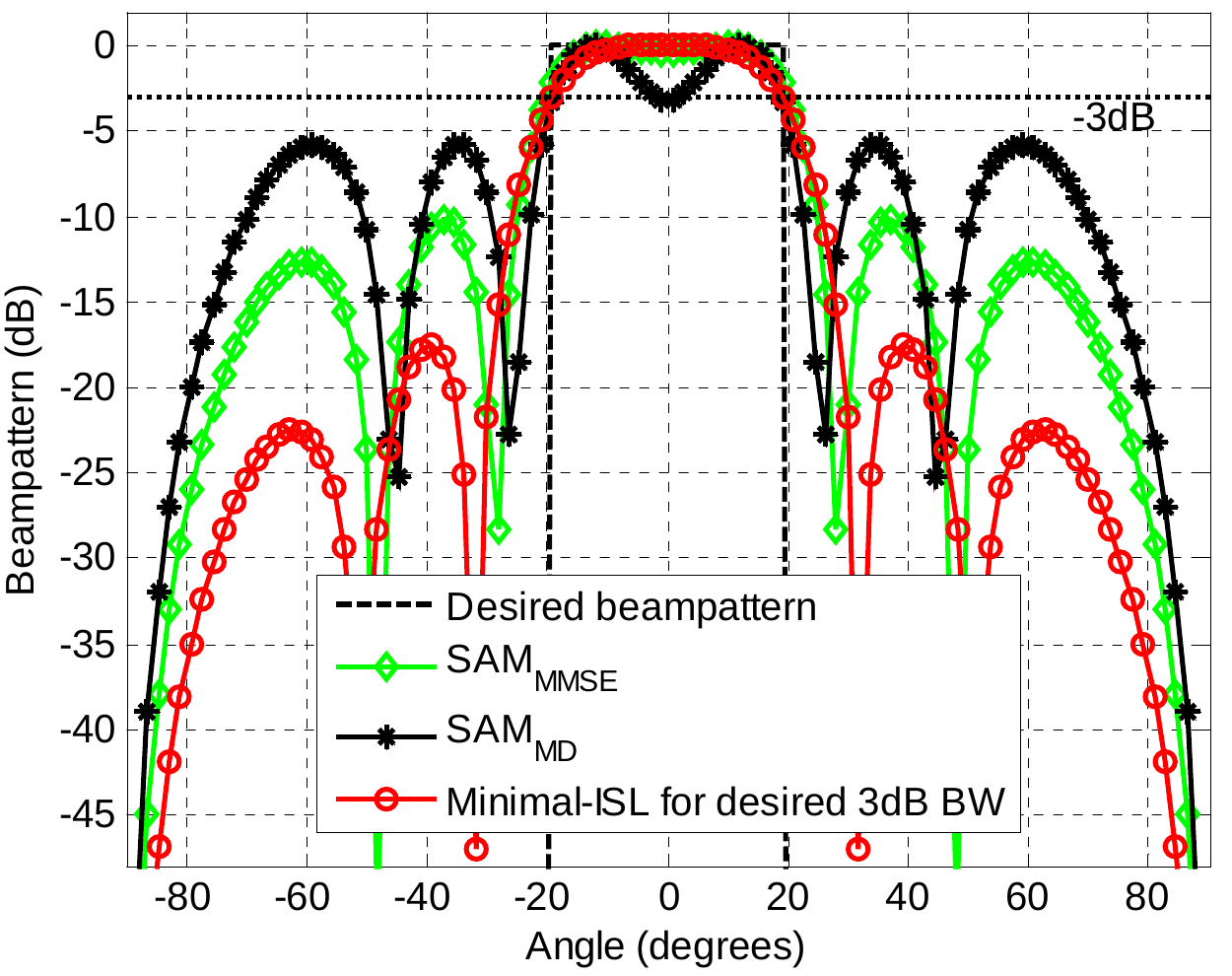}\label{subfig:5b}
  \end{minipage}}\\
  \caption{Beampattern comparisons between the proposed methods and conventional methods.
   (a) Comparison between minimal-ISL only design and DPSSD, (b) Comparison between minimum ISL for a 3dB beamwidth and SAMs.}\label{fig:5_BPCompar2}
\end{figure}
\subsubsection{Comparisons with conventional methods} The minimal-ISL only design method is  compared  with the DPSSD method proposed in \cite{Hassanien2011}, since both of them can obtain closed-form solutions. Fig. \ref{subfig:5a} shows the numerical results of the two methods (where $M=8$, $Q=2$, and the width of the defined mainlobe region is $22^{\circ}$). We can see that our method obtains a lower sidelobe level, which means the closed-formed minimal solution of our method has better performance in terms of lowering sidelobe level. The minimal-ISL design method for the desired 3dB beamwidth is compared with the shape approximation methods (SAMs) which also provide the desired 3dB beamwidth. Now consider two specific SAM methods.  The first one is the SAM proposed in \cite[Sec. III-C]{Stoica2007} (denoted as '$\mathbf{\mathrm{SAM_{MMSE}}}$'). Though $\mathrm{SAM_{MMSE}}$ is a covariance matrix ($\mathbf{R}$) design method, it can obtain the globally optimal solution by using the criteria of minimum MSE. The second method is the SAM proposed in \cite{Khabbazibasmenj2014} (denoted as '$\mathrm{SAM_{MD}}$'), which used shape approximation by minimizing the maximum difference. Fig. \ref{subfig:5b} shows the simulation results of a desired beampattern
with a $40^{\circ}$ beamwidth, where $M=8$ and $Q=3$. From the figure, we can see that the beampattern designed by our method has the lowest sidelobe level.
\subsubsection{Narrow-beam comparisons with conventional phased-array radar}
\begin{figure}[ht!]
  \subfigure[]
  {\begin{minipage}{1\textwidth}
  \centering
  \includegraphics[scale=0.64]{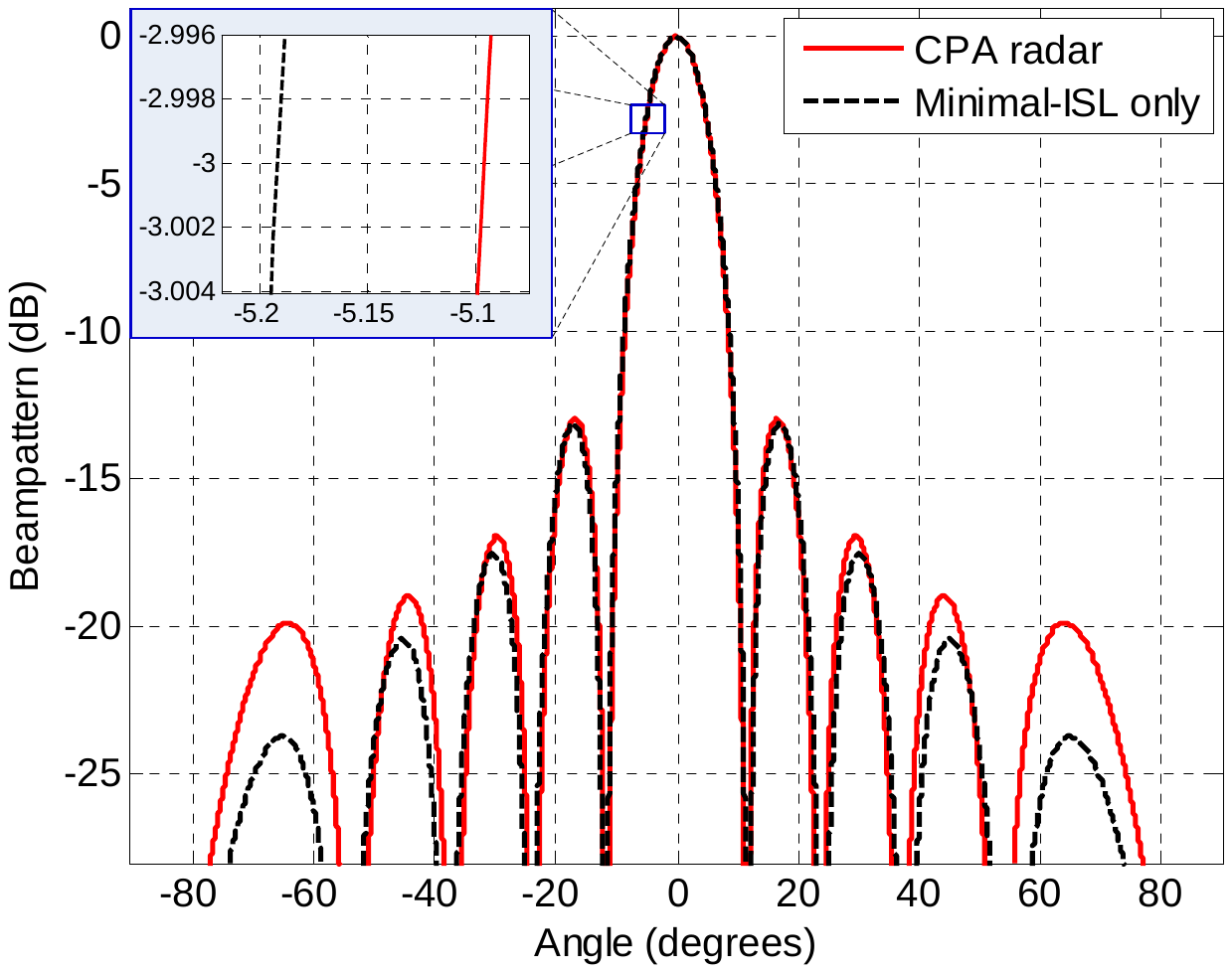}\label{subfig:6a}
  \end{minipage}}\\
  \subfigure[]
  {\begin{minipage}{1\textwidth}
  \centering
  \includegraphics[scale=0.64]{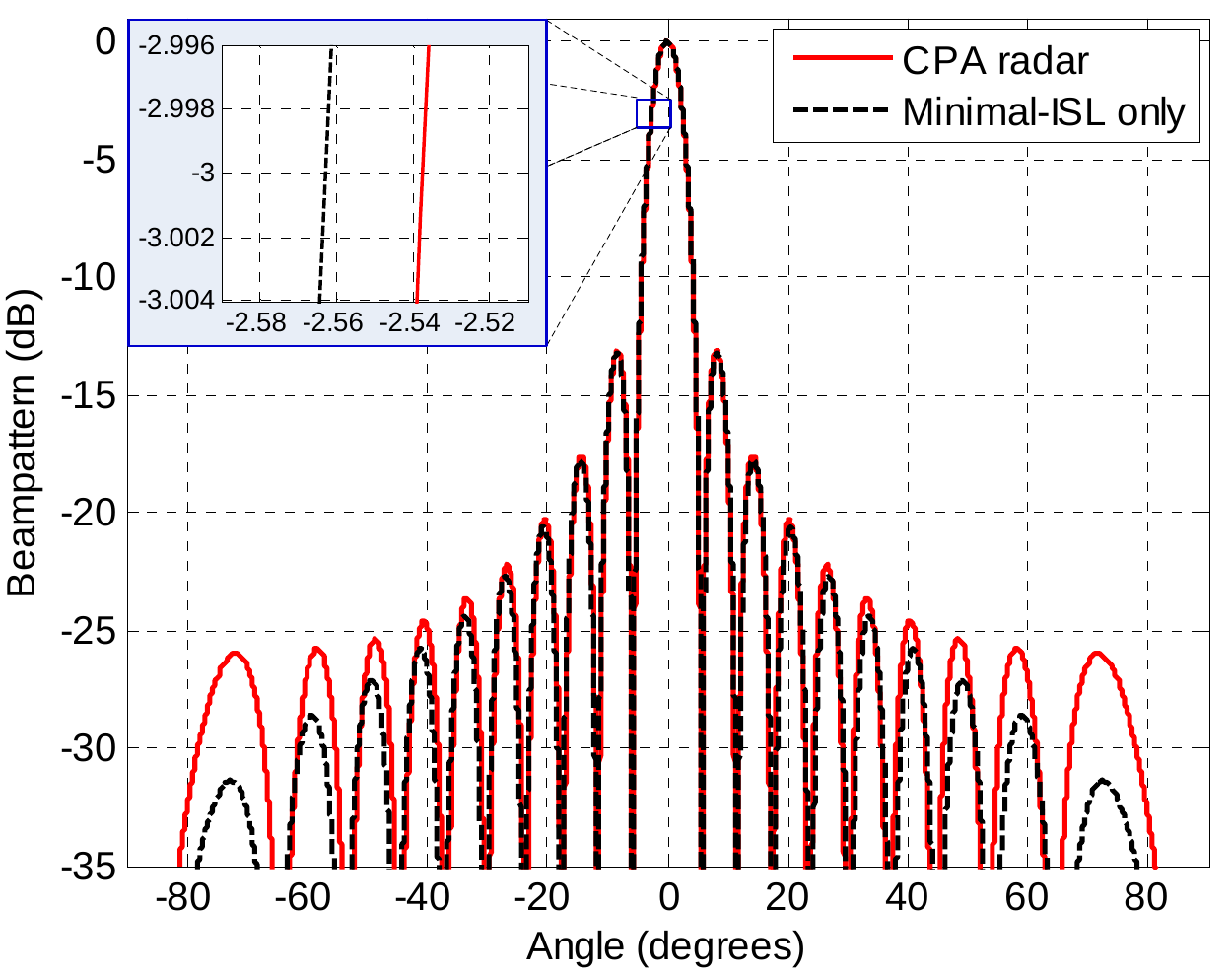}\label{subfig:6b}
  \end{minipage}}\\
  \subfigure[]
  {\begin{minipage}{1\textwidth}
  \centering
  \includegraphics[scale=0.64]{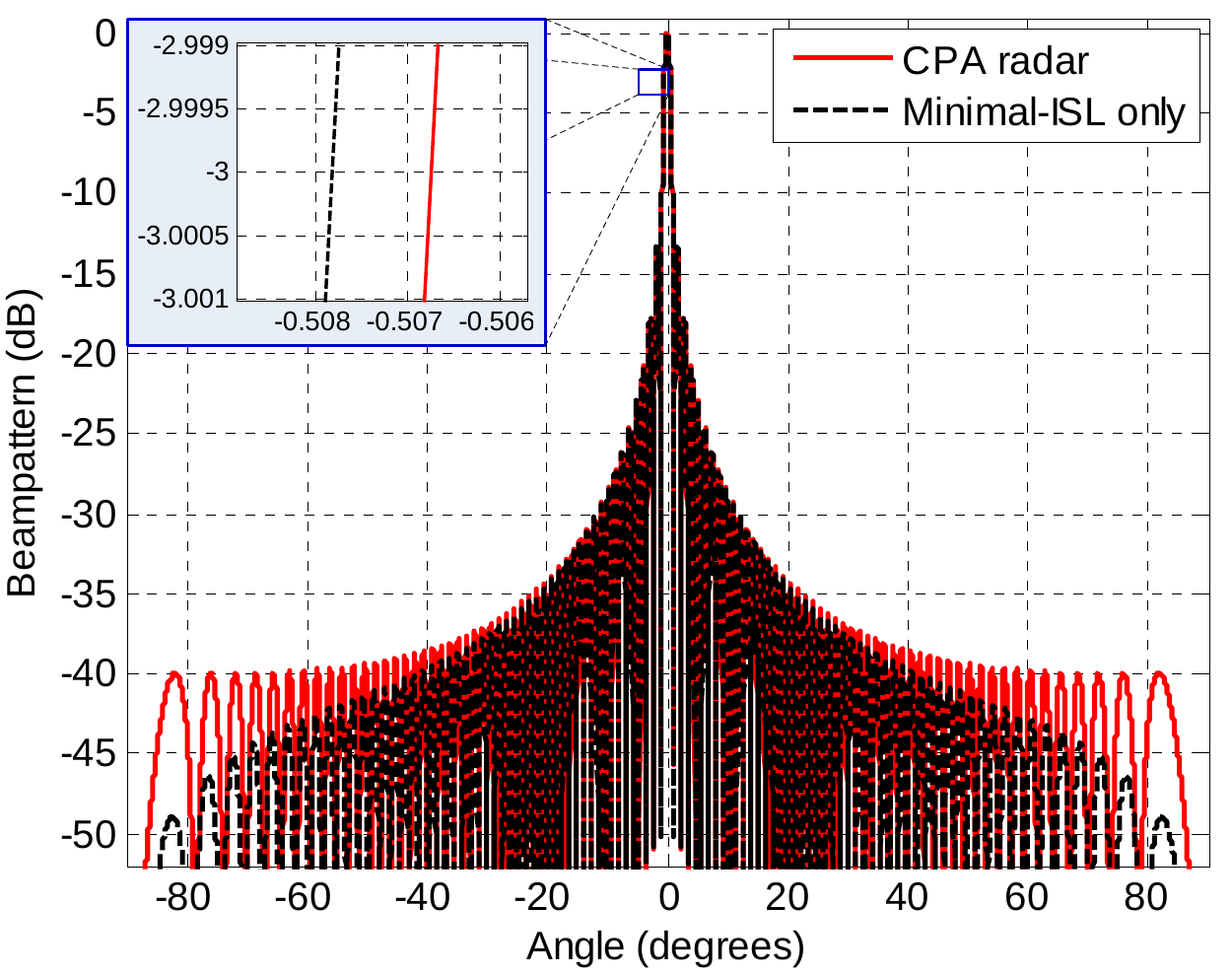}\label{subfig:6c}
  \end{minipage}}\\
  \caption{Single-angle focused beampattern comparisons between the minimal-ISL criterion and CPA radar.
   (a) $M=10$, (b) $M=20$, (c) $M=100$.}\label{fig:6_NBCompar}
\end{figure}

In some cases it is desired to focus energy on a single angle $\theta_{t}$. This can be accomplished by setting $\mathbf{A}_{ml}=\mathbf{I}_{Q}\otimes\left(\mathbf{a}(\theta_{t})\mathbf{a}^{H}(\theta_{t})\right)$ in (\ref{equ:optimization}) and then applying Theorem~1. In Fig.~\ref{fig:6_NBCompar}, we present some comparisons of results obtained from this slight modification of the minimal-ISL only criterion with those obtained from the conventional phased-array (CPA) radar \cite{Skolnik1990}, which can obtain narrowly focused beampatterns \cite{Fuhrmann2004}. The three subfigures in Fig.~\ref{fig:6_NBCompar} show that the beampatterns designed using our method have lower sidelobe levels when compared to those obtained by the CPA radar. When we carefully inspect the 3dB beamwidths of the beampatterns, we can observe that our method can achieve a mainlobe width which seems to approach to that of the conventional one as we increase $M$. As Fig.~\ref{fig:6_NBCompar} illustrates, the differences of the 3dB beamwidths for $M=10$, $M=20$ and $M=100$ are less than $0.1^{\circ}$, $0.02^{\circ}$ and $0.0012^{\circ}$, respectively.
\section{Conclusion}\label{sec:Conclusion}
 \par Colocated MIMO radar waveform design for transmit beampattern formation by minimizing ISL has been considered in this paper. Both analytical and convex design methods using the criteria of minimum ISL are proposed, respectively, to obtain beampatterns with lower sidelobe levels with different goals. Under the minimum ISL design criterion, both theoretical and numerical analyses have shown that the number of waveforms $Q$ doesn't have impact on the quality of beampatterns while the number of transmitters $M$ does. Further, the larger the value of $M$, the lower the value of the sidelobe level. Finally, numerical comparisons have shown our methods can obtain beampatterns with lower sidelobe levels than conventional methods.
\appendices
\section{Proof of Lemma 2}\label{App:ProofLemma2}
\par Define $k_{\star}$ and $\mathbf{X}_{\star}\succeq 0$ such that $k_{\star}=\frac{\mathrm{tr}\{\mathbf{B}\mathbf{X}_{\star}\}}{\mathrm{tr}\{\mathbf{A}\mathbf{X}_{\star}\}}$.
Use eigen-decomposition to factorize $\mathbf{X}_{\star}$ as $\mathbf{X}_{\star}=\sum_{i=1}^{n}\mathbf{x}_{i}\mathbf{x}_{i}^{H}$. Without loss of generality, we assume $\frac{\mathrm{tr}\{\mathbf{B}\mathbf{x}_{1}\mathbf{x}_{1}^{H}\}}
{\mathrm{tr}\{\mathbf{A}\mathbf{x}_{1}\mathbf{x}_{1}^{H}\}}\leq\frac{\mathrm{tr}\{\mathbf{B}\mathbf{x}_{i}\mathbf{x}_{i}^{H}\}}
{\mathrm{tr}\{\mathbf{A}\mathbf{x}_{i}\mathbf{x}_{i}^{H}\}}$ for $i=1,...,n$ and $\mathbf{r}_{min}$ is the minimal solution of $\min\limits_{\mathbf{r}\in {\mathbf{\mathbb{C}}^{n\times1}}}\frac{{{\mathbf{r}^H}{\mathbf{B}}\mathbf{r}}}{{{\mathbf{r}^H}{\mathbf{A}}\mathbf{r}}}$, then
\begin{equation}\label{equ:AppendixProLemma2_1}
\begin{split}
    k_{\star}&=\frac{\mathrm{tr}\{\mathbf{B}\mathbf{X}_{\star}\}}
{\mathrm{tr}\{\mathbf{A}\mathbf{X}_{\star}\}}=\frac{\sum_{i=1}^{n}{\mathrm{tr}\{\mathbf{B}\mathbf{x}_{i}\mathbf{x}_{i}^{H}\}}}
{\sum_{i=1}^{n}{\mathrm{tr}\{\mathbf{A}\mathbf{x}_{i}\mathbf{x}_{i}^{H}\}}}\\
&\geq \min\limits_{i}\frac{\mathrm{tr}\{\mathbf{B}\mathbf{x}_{i}\mathbf{x}_{i}^{H}\}}{\mathrm{tr}\{\mathbf{A}\mathbf{x}_{i}\mathbf{x}_{i}^{H}\}}
=\frac{\mathrm{tr}\{\mathbf{B}\mathbf{x}_{1}\mathbf{x}_{1}^{H}\}}
{\mathrm{tr}\{\mathbf{A}\mathbf{x}_{1}\mathbf{x}_{1}^{H}\}}\\
&\geq\frac{\mathrm{tr}\{\mathbf{B}\mathbf{r}_{min}\mathbf{r}_{min}^{H}\}}
{\mathrm{tr}\{\mathbf{A}\mathbf{r}_{min}\mathbf{r}_{min}^{H}\}}=k_{min}
\end{split}
\end{equation}
\par Note that $\mathbf{X}\star$ can be an arbitrary feasible point of $\min\limits_{\mathbf{X}\succeq 0} \frac{\mathrm{tr}\{\mathbf{B}\mathbf{X}\}}{\mathrm{tr}\{\mathbf{A}\mathbf{X}\}}$, thus $\min\limits_{\mathbf{X}\succeq 0} \frac{\mathrm{tr}\{\mathbf{B}\mathbf{X}\}}{\mathrm{tr}\{\mathbf{A}\mathbf{X}\}}\geq k_{min}$. On the other hand, we have $\min\limits_{\mathbf{X}\succeq 0} \frac{\mathrm{tr}\{\mathbf{B}\mathbf{X}\}}{\mathrm{tr}\{\mathbf{A}\mathbf{X}\}} \leq k_{min}$ for $\min\limits_{\mathbf{X}\succeq 0} \frac{\mathrm{tr}\{\mathbf{B}\mathbf{X}\}}{\mathrm{tr}\{\mathbf{A}\mathbf{X}\}}$ is an SDR of $\min\limits_{\mathbf{r}\in\mathbf{\mathbb{C}}^{n\times1}}\frac{{{\mathbf{r}^H}{\mathbf{B}}\mathbf{r}}}{{{\mathbf{r}^H}{\mathbf{A}}\mathbf{r}}}$. Therefore the proof of Lemma 2 is completed. $\hfill\blacksquare$
\section{Proof of Lemma 3}\label{App:ProofLemma3}
\par Write $f(Q+1)$ as $f(Q+1)=\min\limits_{\mathbf{X}_{+}\succeq 0}{\mathrm{tr}\left\{ {\left( {{\mathbf{I}_{Q+1}} \otimes \mathbf{B}} \right)\mathbf{X}_{+}} \right\}}$ $\mathrm{s.t.}$ ${\mathrm{tr}\left\{ {\left( {{\mathbf{I}_{Q+1}} \otimes \mathbf{A}_{i}} \right)\mathbf{X}_{+}} \right\}}\leq b_{i},~i=1,...,K$, since if we let $\mathbf{X}_{+}=\left(\begin{array}{cc}
                                                                                         \mathbf{X} & \mathbf{0} \\
                                                                                         \mathbf{0} & \mathbf{0}_{M\times M}
                                                                                       \end{array}\right)(\in\mathbf{\mathbb{C}}^{M(Q+1)\times M(Q+1)})
$ then $f(Q+1)=f(Q)$ holds, hence, we have
 \begin{equation}\label{equ:ProofLemma31}
   f(Q+1)\leq f(Q)
 \end{equation}
 \par Define $\mathbf{X}_{+}=\left(\begin{array}{cc}
                                      \mathbf{X}_{1} & \mathbf{X}_{12} \\
                                      \mathbf{X}_{12}^{H} & \mathbf{X}_{2}
                                    \end{array}
 \right)$, where $\mathbf{X}_{1}\in\mathbf{\mathbb{C}}^{MQ\times MQ}$, $\mathbf{X}_{12}\in\mathbf{\mathbb{C}}^{MQ\times M}$ and $\mathbf{X}_{2}\in\mathbf{\mathbb{C}}^{M\times M}$, then we can rewrite $f(Q+1)$ as
 \begin{equation}\label{equ:ProofLemma32}
 \begin{split}
   f(Q+1)=&\min\limits_{\mathbf{X}_{+}\succeq 0}{\mathrm{tr}\left\{ {\left( {{\mathbf{I}_{Q}} \otimes \mathbf{B}} \right)\mathbf{X}_{1}}\right\}+\mathrm{tr}\left\{{\mathbf{B} \mathbf{X}_{2}} \right\}}~~\mathrm{s.t.}\\
\mathrm{tr}&{\left\{ {\left( {{\mathbf{I}_{Q}} \otimes \mathbf{A}_{i}} \right)\mathbf{X}_{1}}\right\}+\mathrm{tr}\left\{{{\mathbf{A}_{i}} \mathbf{X}_{2}} \right\}}\leq b_{i},~i=1,...,K\\
     =&\min\limits_{\mathbf{X}_{1},\widetilde{\mathbf{X}}_{2}\succeq 0}{\mathrm{tr}\left\{ {\left( {{\mathbf{I}_{Q}} \otimes \mathbf{B}} \right)\left(\mathbf{X}_{1}+\widetilde{\mathbf{X}}_{2}\right)} \right\}}~~\mathrm{s.t.}\\
\mathrm{tr}&{\left\{ {\left( {{\mathbf{I}_{Q}} \otimes \mathbf{A}_{i}} \right)\left(\mathbf{X}_{1}+\widetilde{\mathbf{X}}_{2}\right)} \right\}}\leq b_{i},~i=1,...,K
 \end{split}
 \end{equation}
 where $\widetilde{\mathbf{X}}_{2}\in\mathbf{\mathbb{C}}^{MQ\times MQ}$ and $\widetilde{\mathbf{X}}_{2}=\left(\begin{array}{cc}
                                           \mathbf{X}_{2} & \mathbf{0} \\
                                           \mathbf{0} & \mathbf{0}_{M(Q-1)\times M(Q-1)}
                                         \end{array}
 \right)$.
 \par Obviously, the minimization in $f(Q+1)$ (see the RHS of (\ref{equ:ProofLemma32})) can be relaxed to the minimization in $f(Q)$, thus we have $f(Q+1)\geq f(Q)$. Combining (\ref{equ:ProofLemma31}) then we have $f(Q+1)=f(Q)$. Therefore, the proof of Lemma 3 is completed.$\hfill\blacksquare$

\ifCLASSOPTIONcaptionsoff
  \newpage
\fi

%
\end{document}